\documentclass[12pt, a4paper]{article}
\pdfoutput=1

\usepackage{amsmath}
\usepackage{amsfonts}
\usepackage{amssymb}
\usepackage{graphicx, rotating}
\usepackage{epsfig}
\usepackage{latexsym}
\usepackage{graphicx}
\usepackage{color}
\usepackage{amsmath,bm,amssymb}
\usepackage{cite}
\usepackage{slashed}
\usepackage{hyperref}
\hypersetup{colorlinks, citecolor=bluscuro, linkcolor=black, urlcolor=bluscuro}
\definecolor{rossos}{cmyk}{0,1,1,0.55}
\definecolor{bluscuro}{rgb}{0.15, 0.2, .85}
\definecolor{bluchiaro}{cmyk}{1,.3,0.,0.1}

\numberwithin{equation}{section}

\newcommand{\be}{\begin{equation}}
\newcommand{\ee}{\end{equation}}
\newcommand{\bea}{\begin{eqnarray}}
\newcommand{\eea}{\end{eqnarray}}

\def\simlt{\stackrel{<}{{}_\sim}}

\def\pp{{\scriptscriptstyle +}}
\def\mm{{\scriptscriptstyle -}}
\def\pms{{\scriptscriptstyle{\pm}}}

\newcommand{\arXiv}[2]{\href{http://arxiv.org/pdf/#1}{{\tt [#2/#1]}}}
\newcommand{\arXivold}[1]{\href{http://arxiv.org/pdf/hep-#1}{{\tt [#1]}}}

\def\bma#1{\mbox{\boldmath{$#1$}}}

\begin{document}
\allowdisplaybreaks
\begin{titlepage}
\begin{flushright}
IFT-UAM/CSIC-20-72
\end{flushright}
\vspace{.3in}

\vspace{1cm}
\begin{center}
{\Large\bf\color{black} 
Exactly Solvable Vacuum Decays with Gravity}
\\
\vspace{1cm}{
{\large J.R.~Espinosa$^1$, J.-F. Fortin$^2$} and {\large J. Huertas$^1$}
} \\[7mm]
{\it ${}^1$ Instituto de F\'{\i}sica Te\'orica UAM/CSIC, \\ 
C/ Nicol\'as Cabrera 13-15, Campus de Cantoblanco, 28049, Madrid, Spain\\
${}^2$ D\'epartement de Physique, de G\'enie Physique et d'Optique\\
Universit\'e Laval, Qu\'ebec, QC G1V 0A6, Canada
}

\end{center}
\bigskip

\vspace{.4cm}

\begin{abstract}
Using a new approach to the analysis of false vacuum decay based on the so-called tunneling potential,
 we develop a general method to find scalar potentials with a false vacuum with exactly solvable decay at the semi-classical level, including gravitational corrections. We examine in particular the decays of de Sitter vacua providing concrete  examples that allow to explore analytically the transition between the Coleman-De Luccia and Hawking-Moss regimes.  
\end{abstract}
\bigskip

\end{titlepage}

\section{Introduction}
False vacuum decay in quantum field theory is a fascinating phenomenon of key importance for cosmology, both in the Standard Model and its extensions. A giant step in the theoretical analysis of such decays (in particular for the calculation of the tunneling action that controls the exponential suppression of vacuum decay) was due to Coleman \cite{Coleman} who pioneered an elegant and powerful approach 
based on an Euclidean formulation of the problem. This formulation is well known, as it became the standard approach, it proved particularly convenient to go beyond the semi-classical approximation including quantum corrections \cite{CC}, and was also extended, by Coleman and De Luccia, to include gravitational corrections \cite{CdL}.

Recently, an alternative approach to the calculation of tunneling actions was proposed \cite{E}. It reformulates the problem without reference to Euclidean space or quantities, as a variational problem in field space, using an auxiliary function, $V_t$, dubbed tunneling potential, that contains all the information needed to calculate the decay action. It is always useful to have alternative formulations for important problems, like false vacuum decay certainly is, and the $V_t$ approach has proven quite useful in many respects, see \cite{Eg,EK,EnoB,ESM,Estab,RM}. 

It was also possible to extend the $V_t$ formalism to include in a very compact way gravitational corrections \cite{Eg}.  In this paper we  make extensive use of this general formulation, reviewed in section~\ref{sec:Vt} and push forward in several directions the results obtained already in \cite{Eg}. One particularly interesting type of vacuum decay is that of de Sitter (dS) vacua. While in \cite{Eg} it was already discussed how the $V_t$ formalism recovered the Hawking-Moss rate \cite{HM} in the appropriate limit, we enlarge that discussion here, explaining in more detail how this new formalism reproduces the detailed balance for transitions between dS vacua (section~\ref{sec:dStodS}), some basic properties of the dS decay action under several rescalings of parameters (section~\ref{sec:basics}) and different ways in which
the Hawking-Moss (HM) regime takes over the usual Coleman-De Luccia (CdL) one (section~\ref{sec:HM}), for which the $V_t$ approach is ideally suited.

While the CdL to HM transition is discussed in an approximate way in section~\ref{sec:HM}, one of the useful applications of the $V_t$ approach is to find potentials, $V$, for which the problem of finding the decay configuration is exactly solvable. In \cite{E} this was done in a generic way formally integrating an equation for $V$ in terms of $V_t$, that can be explicitly solved for simple $V_t$ choices. With gravity included, this problem is harder and \cite{Eg} did not provide a general solution, although it did
manage to give an exactly solvable $V$ for a particular $V_t$. This  solvable model, which had some adjustable parameters, could provide analytic examples for decays of Minkowski and anti-de Sitter (AdS) false vacua. No example was given for dS false vacua and the possibility of analytically examining the switch from a CdL-dominated decay to a HM-dominated one was not achieved. 
In this paper we fix both shortcomings of \cite{Eg}: in section~\ref{sec:exact} we develop a general method to find exactly solvable potentials if $V_t$ is given, including gravity. The method is applicable to any false vacua, Minkowski, dS or AdS. In section~\ref{sec:ex} we collect examples of solvable potentials with false dS vacua decaying to other dS  or  AdS vacua. This serves, in particular, to study in an analytically controllable way how CdL-dominated decays turn to HM-dominated transitions.  We conclude in section~\ref{sec:conclusions}.

\section{Brief Review of the Tunneling Potential Approach\label{sec:Vt}}

The so-called tunneling potential approach \cite{E,Eg} is an alternative formulation of the problem of finding the tunneling action for the decay of a false vacuum at $\phi_\pp$ of some potential $V(\phi)$. It does not rely on the Euclidean
formulation of Coleman \cite{Coleman} and transforms the problem into a simple variational problem in field space. In the case with gravity, this variational problem is the following: find the (tunneling potential) function $V_t(\phi)$ which interpolates between $\phi_\pp$ and some unknown $\phi_0$ on the basin of the true vacuum\footnote{If we call $\phi_\mm$ the true vacuum, we use the convention $\phi_\mm>\phi_\pp$. Then, $\phi_\pp<\phi_0< \phi_\mm$.} and minimizes the action functional
\be
 S[V_t]=\frac{6\pi^2}{\kappa^2}\int_{\phi_\pp}^{\phi_0}d\phi\ \frac{(D+V_t')^2}{V_t^2 D}\ ,
\label{SVt}
\ee
with 
\be
D^2\equiv V_t'{}^2+6\kappa (V-V_t)V_t\ ,
\label{D2}
\ee
where primes denote field derivatives, and $\kappa=1/m_P^2$  ($m_P$ is the reduced Planck mass). This method reproduces the Euclidean result of Coleman and De Luccia \cite{CdL} and has a number of advantages discussed elsewhere \cite{E,Eg,Estab,EK}. 
The Euler-Lagrange equation $\delta S/\delta V_t=0$  gives 
\be
\frac{\delta S}{\delta V_t} = -108 \pi^2 \frac{(V-V_t)}{D^5}\  {\rm{EoM}}=0\ ,
\label{EoM}
\ee
where the  ``equation of motion'' (EoM) for $V_t$ is:
\be
{\rm{EoM}}\equiv 
6(V-V_t)\left[V_t''+\kappa\left(
3V-2V_t\right)\right]
+ V_t'\left(4V_t'-3V'\right)=0
\ ,
\label{EoMVt}
\ee
or, in terms of $V_t'/D$,
\be
\frac{d}{d\phi}\left(\frac{V_t'}{D}\right)= \kappa\frac{\left(2V_t-3V  \right)}{D}\ .
\label{EoMVtD}
\ee

The tunneling potential $V_t$ satisfies the boundary conditions 
\be
V_t(\phi_+)=V(\phi_+)\equiv V_\pp\ ,\quad\quad  V_t(\phi_0)=V(\phi_0)\equiv V_0\ ,
\label{BCsVt}
\ee
where $\phi_0$ must be determined by minimizing (\ref{SVt}) 
and equals the central value of the bounce, $\phi(0)$, in the Euclidean approach. 

\begin{figure}[t!]
\begin{center}
\includegraphics[width=0.5\textwidth]{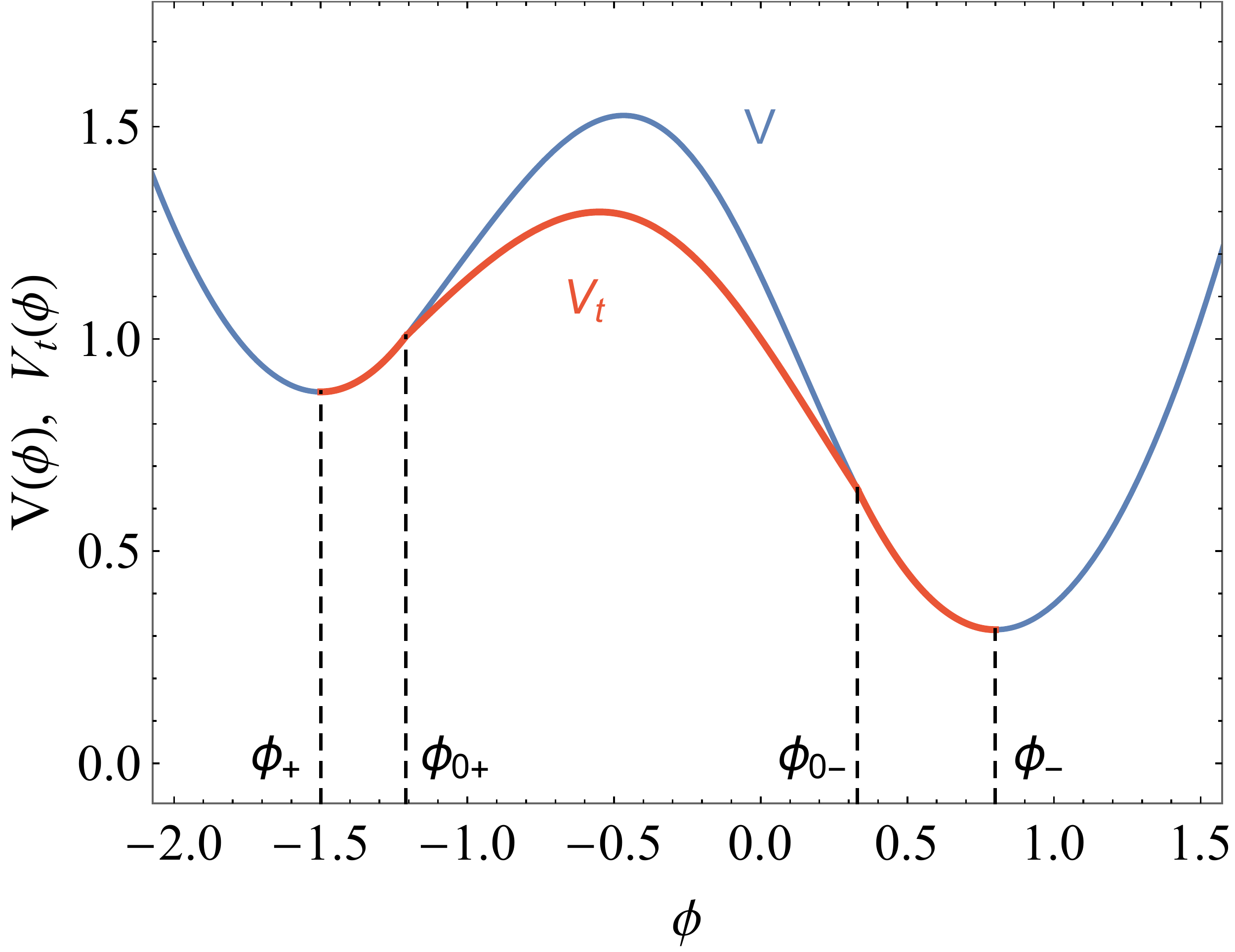}
\end{center}
\vspace{-0.5cm}
\caption{Potential $V(\phi)$ with two dS mimima at $\phi_\pm$ and
tunneling potential $V_t(\phi)$ describing the transitions between them. The interval $(\phi_{0+},\phi_{0-})$ is the field range of the  CdL bounce. 
\label{fig:dStodS}
}
\end{figure}

$V_t$ is qualitatively different depending on the false vacuum nature \cite{Eg} :
\begin{itemize}

\item For decays of Minkowski or AdS false vacua, $V_t$ is monotonic with $V_t'\leq 0$. The EoM for $V_t$ also fixes
$V_t'(\phi_+)=V'(\phi_+)=0,$ and $V_t'(\phi_0)=3 V'(\phi_0)/4.$
As known from \cite{CdL}, for this type of false vacua, gravity can forbid decay (gravitational quenching). In the $V_t$ formulation, to have
a real tunneling action, $V_t$ must  give $D^2>0$.
Gravitational quenching occurs when this condition cannot be satisfied for any $V_t$ \cite{Eg}. For these vacua the second term in (\ref{D2}) is negative and in some cases it can be impossible to satisfy  $D^2>0$ for any $V_t$ and the potential is stabilized \cite{Estab}.

\item For decays of dS  vacua, $V_t$ is not monotonic, see figure~\ref{fig:dStodS}. From $\phi_+$ to $\phi_{0+}$ one has $V_t=V$
[notice that this also solves $\delta S/\delta\phi=0$, see (\ref{EoM})].  From $\phi_{0+}$ to some $\phi_{0-}$, $V_t<V$  and this range corresponds to the field range for the CdL bounce, with
$V_t'(\phi_{0\pm})=3V'(\phi_{0\pm})/4$ and $\phi_{0-}$ corresponding to $\phi(0)$ of the bounce. Finally, from $\phi_{0-}$ to the second minimum $\phi_-$, one has again $V_t=V$. If the overall energy scale of the potential is increased, the range of the CdL interval shrinks to zero and the action tends to the HM one \cite{Eg}.
 
\end{itemize}

For later reference we collect here several formulas that connect the tunneling potential formalism to the Euclidean approach and allow to translate the results between the two. For details on their derivation see \cite{Eg}. In the Euclidean formulation, vacuum decay is described by a bounce configuration $\phi(\xi)$, an $O(4)$-symmetric  extremal of the Euclidean action, and a metric function, $\rho(\xi)$, for the $O(4)$-symmetric  Euclidean space-time metric 
\be
ds^2= d\xi^2 +\rho(\xi)^2 d\Omega_3^2\ ,
\ee
where $\xi$ is a radial coordinate and $d\Omega_3^2$ is the line element on a unit three-sphere. 

The key link between both formulations is
\be
V_t (\phi)= V(\phi) -\frac12 \dot\phi^2\ ,
\label{Vtdef}
\ee
where $\dot x\equiv dx/d\xi$, and $\dot\phi$ on the right hand side is assumed to be expressed in terms of the field, using the bounce profile $\phi(\xi)$. Using (\ref{Vtdef}) and the differential equations satisfied by $\phi$ and $\rho$ one can establish the following dictionary, where the left hand side is a Euclidean quantity and the right hand side does not depend at all on Euclidean quantities. We have
\be
\dot\phi = - \sqrt{2(V-V_t)}\ ,\quad \ddot\phi=V'-V_t'\ ,
\label{dphi}
\ee
where the minus sign in the equation for $\dot\phi$ applies for our convention $\phi_\pp<\phi_\mm$, and
\be
\rho=\frac{3\sqrt{2(V-V_t)}}{D}\ ,\quad 
\dot\rho =-\frac{V_t'}{D}\ ,\quad 
\frac{\ddot\rho}{\rho} = -\frac{\kappa}{3}\ (3V-2V_t)\ .
\label{rho}
\ee

\section{dS to dS Transitions\label{sec:dStodS}}

In the decay of a dS vacuum, only the finite space inside the horizon is required to transition (see e.g. \cite{Brown}). This makes the rate non zero generically and allows upward transitions, from a dS vacuum to another with higher cosmological constant. There is a simple relation between the actions for transitions between two dS vacua and it is instructive to derive such relation in the tunneling potential formalism.

The tunneling action $S_{\pp\to\mm}$ for the decay from a dS vacuum at $\phi_\pp$ to a dS vacuum at $\phi_\mm$ can be obtained as the action integral 
(\ref{SVt}). For the discussion in this section it is convenient to extend the integration interval to the full interval from $\phi_\pp$ to $\phi_\mm$. The integral has three different pieces \cite{Eg} according to the three different ranges described in the previous section and illustrated by figure~\ref{fig:dStodS}: In the first, from $\phi_\pp$ to $\phi_{0\pp}$, where one has $V_t\equiv V$, with $V_t'=V'\geq 0$, one gets $D=V_t'$ and the action density is simply
\be
{\it s}=\frac{24\pi^2 V'}{\kappa^2 V^2}
\label{strivial}
\ ,
\ee
which can be integrated exactly.
 In the CdL range, $(\phi_{0\pp},\phi_{0\mm})$, one has $V_t\leq V$. Finally, from $\phi_{0\mm}$ to $\phi_{\mm}$ with $V_t\equiv V$ again  but with $V_t'=V'\leq 0$, the action density is $s=0$.\footnote{ One also sees that there are no transitions from AdS to dS: the action density would diverge if $V_\pp<0$ as $V_t$ would need to cross zero with positive slope.}  The total decay action can then be written as the sum of two nonzero pieces, one that we call (with an slight abuse of notation) the ``Hawking-Moss''  part and the other the CdL contribution\footnote{The mixed nature of the dS to dS transitions, with a ``HM'' piece and a CdL part has been given a thermal interpretation in \cite{BW}.}:
\be
S_{\pp\to\mm}= \Delta S_{HM}+\Delta S_{CdL}=\frac{24\pi^2}{\kappa^2}\left(\frac{1}{V_\pp}-\frac{1}{V_{0\pp}}\right)+\frac{6\pi^2}{\kappa^2}\int_{\phi_{0\pp}}^{\phi_{0\mm}}\frac{(D+V_t')^2}{D V_t^2}\ d\phi\ ,
\label{SHMCdL}
\ee
with $V_{0\pp}\equiv V(\phi_{0\pp})$. When the overall mass scale of the potential (say $V_\pp$) is sufficiently large, the CdL bounce part disappears and the transition is purely of Hawking-Moss type, as discussed in section~\ref{sec:HM}.

The decay in the opposite direction, from $\phi_\mm$ to $\phi_\pp$ proceeds in a similar manner, in fact with the same $V_t$ function, but now taken as starting from $\phi_\mm$, so that its derivative flips sign. This implies that now there is a simple non-zero contribution from the interval $\phi_\mm$  to $\phi_{0\mm}$ and a zero contribution from the interval from $\phi_{0\pp}$ to $\phi_\pp$.
The difference between the two tunneling actions, $\Delta S\equiv S_{\pp\to\mm}-S_{\mm\to\pp}$, takes a very simple form, as only the term linear in $V_t'$ in the action density, the only one that flips sign, contributes. This term can be integrated exactly and one gets
\be
\Delta S =\frac{24\pi^2}{\kappa^2}\left(\frac{1}{V_\pp}-\frac{1}{V_\mm}\right)\ .
\ee 
This can be rewritten simply as $\Delta S = S_\pp - S_\mm$,
where $S_\pm$ is the Gibbons-Hawking entropy of a dS vacuum with cosmological constant $V_\pm$. Indeed, this entropy is one fourth of the horizon's area in Planck units
$
S_\pm=A_\pm/(4l_P^{2})$,
where the area is given by
$
A_\pm=4\pi/ H_\pm^{2}\ ,
$
with $H_\pm^2=\kappa V_\pm/3$.
In the formulas above one has $l_P=1/M_P$ and $1/M_P^2=G$, $8\pi G=\kappa$.

\section{Some Basic Properties of dS Decays\label{sec:basics}}
The expression for the tunneling action in terms of $V_t$ can be used to derive in a simple way the following basic properties for dS vacuum decays under several rescalings:  (1) Stretching of the potential; (2) stretching of the field; (3) changing the strength of gravity and (4) a combination of the latter two. 

(1) If the potential is rescaled by a constant, $V\to a V$, then the tunneling potential that minimizes the action is $a V_t$ (as is clear from the EoM for $V_t$) and the tunneling action gets rescaled as
$S\to S/a$. Therefore, for $a>1$ the rescaling of $V$ increases the energy scale of the false vacuum and makes it more unstable, even though the height of the potential barrier is also increased.  This conclusion holds whether the transition happens via a CdL bounce or via a HM instanton.

(2) The effect of gravity can be examined by studying how the action is affected by a change in $\kappa$ [in the understanding that a larger (smaller) $\kappa$ means all mass scales of the potential are smaller (larger) compared to $m_P$]. This is best analyzed by looking at 
\be
\frac{dS[V_t]}{d\kappa}= \frac{d}{d\kappa}\int_{\phi_\pp}^{\phi_\mm}s(V,V_t,V_t',\kappa)d\phi =\int_{\phi_\pp}^{\phi_\mm}\left[\left(\frac{\partial s}{\partial V_t}-\frac{d}{d\phi}\frac{\partial s}{\partial V_t'}\right)\frac{dV_t}{d\kappa}+\frac{\partial s}{\partial \kappa}\right]d\phi
\ ,
\ee
where $s(V,V_t,V_t',\kappa)$ is the tunneling action density and we have written the total action as an integral in the full interval $(\phi_\pp,\phi_\mm)$, as in the previous section.
The term inside round brackets above cancels due to the EoM for $V_t$, eqs.~(\ref{EoM}) and (\ref{EoMVt}) in the whole integration interval,
and the rest gives
\be
\frac{dS[V_t]}{d\kappa}=\frac{3\pi^2}{\kappa^3} \int_{\phi_\pp}^{\phi_\mm}\frac{D}{V_t^2}\left(1+\frac{V_t'}{D}\right)^2\left[\left(1-\frac{V_t'}{D}\right)^2-4\right]d\phi\ .
\label{dSdk}
\ee
The sign of this derivative depends on the type of transition considered. For dS to dS transitions, defined as those for which $V(\phi_{0\mm})>0$ even if $V_\mm$ might be negative, we have $V_t>0$ for $\phi<\phi_{0\mm}$ (and zero integrand for $\phi>\phi_{0\mm}$). Then, it follows that $-1\leq V_t'/D \leq 1$ and
$dS[V_t]/d\kappa<0$ and so, one concludes that larger $\kappa$ (stronger gravitational effects) lowers the tunneling action, making dS vacua more unstable. 

However, for dS to AdS transitions [i.e. those with $V(\phi_{0\mm})<0$], $V_t$ gets negative at some point resulting in $V_t'/D<-1$ and the sign of
$dS[V_t]/d\kappa$ depends on the shape of the potential.  If the region with negative $V_t$ dominates in the integral (\ref{dSdk}) one would get a positive slope and stronger gravity makes the false vacuum more stable (as always happens for the decays of Minkowski or AdS false vacua). In subsection~\ref{7.5} we show an example of potential that can realize both signs of $dS[V_t]/d\kappa$ depending on its parameters.

(3) If we rescale the field as $\phi\to \phi/a$, the tunneling potential 
for $V_a(\phi)\equiv V(\phi/a)$ is not simply $V_t(\phi/a)$ but we can still see how the action changes depending on whether $a<1$ (thinner barrier) or $a>1$ (wider barrier). As the HM part of the action does not change under a field rescaling, we simply look at the CdL part. Consider the CdL part of the action for a rescaled $V_{ta}(\phi)\equiv V_t(\phi/a)$ (for any arbitrary $V_t$):
\be
\Delta S_{CdL}[V_{ta}]=\frac{6\pi^2}{\kappa^2}\int_{a\phi_{0\pp}}^{a\phi_{0\mm}}\frac{[D^a_\kappa+V_{ta}']^2}{D^a_\kappa V_{ta}^2}d\phi\ ,
\ee
where 
\be
D_\kappa^a\equiv\left[V_{ta}'{}^2+6\kappa(V_a-V_{ta})V_{ta}\right]^{1/2}\ .
\ee
Changing the integration variable $\phi\to \phi/a$ we have
\be
\Delta S_{CdL}[V_{ta}]=\frac{6\pi^2}{\kappa^2}\int_{\phi_{0\pp}}^{\phi_{0\mm}}\frac{[D_{\kappa a^2}+V_t']^2}{D_{\kappa a^2} V_{t}^2}d\phi\ ,
\label{Srescaled}
\ee
where 
\be
D_{\kappa a^2}\equiv\left[V_{t}'{}^2+6\kappa a^2(V-V_{t})V_{t}\right]^{1/2}\ .
\ee
Now, for $a>1$ it is easy to show that $[D_{\kappa a^2}+V_t']^2/(D_{\kappa a^2}V_t^2) \geq [D_{\kappa}+V_t']^2/(D_{\kappa}V_t^2)$ so that the action integrand is larger
for $a>1$ than for $a=1$.\footnote{Taking $x=6\kappa(V-V_t)V_t/V_t'{}^2$ reduces the problem to the inequality $f(x a^2)\geq f(x)$
involving the function $f(x)\equiv \sqrt{1+x}+1/\sqrt{1+x}$ for $x\in (-1,\infty)$,  as $D^2>0$ . The equality holds only for $x=0$, which requires $V=V_t$, i.e. a HM transition} This proves that, for $a>1$,  $\Delta S_{CdL}[V_{ta}]>\Delta S_{CdL}[V_t]$ for any $V_t$. Therefore, the minimum of
the functional $S[V_{ta}]$ is larger (or equal\footnote{If the original decay was already a HM one, then the rescaling has no effect on the action and that is why we also include the possible equality here.}) than the minimum of  $S[V_{t}]$ and the tunneling action must increase (or stay constant) when the barrier is made wider ($a>1$).  Eventually, the action for tunneling via a CdL instanton will become larger than the HM one (unchanged by the field rescaling) and for barriers wider than a critical value it is the HM action that controls the decay. Finally, for the opposite case of $a<1$ all inequalities are reversed, so that a  thinner barrier leads to a higher decay rate (lower tunneling action).

(4) From (\ref{Srescaled}) and the previous discussion of the $\kappa$ dependence of the action, we see that the tunneling action after a field
rescaling $\phi\to \phi/a$ is related to the action after $\kappa\to\kappa a^2$ by $S_{\phi\to\phi/a} = a^4  S_{\kappa\to\kappa a^2}$.
We can combine this inequality with $S_{\phi\to\phi/a}\geq S$ for $a>1$ from (3) above (wider barriers increase the action). For dS to dS transitions or dS to AdS transitions with $dS/d\kappa<0$, we also know from (2) above that $S_{\kappa\to\kappa a^2}<S$ for $a>1$ (stronger gravity makes such dS vacua more unstable). Combining
all this we get $S\leq S_{\phi\to\phi/a} = a^4  S_{\kappa\to\kappa a^2}<a^4 S$. Therefore we arrive at the following two inequalities
for $a>1$
\bea
S\leq S_{\phi\to\phi/a}<a^4 S\ , \nonumber\\
a^{-4} S\leq S_{\kappa\to\kappa a^2}< S\ ,
\label{scalingrange}
\eea
that set upper and lower bounds for these rescalings. Notice that this behaviour is quite different from the one for Minkowski or AdS vacuum decay. In these two cases, making the barrier wide enough or increasing sufficiently the strength of gravity could quench completely the decay (leading to $S\to\infty$). The bounds above prevent this from happening in the dS case. After all, a dS vacuum
can always decay via the HM instanton.

\section{Onset of Hawking-Moss Transitions \label{sec:HM}}

It is well known that, if the vacuum energy of a false dS vacuum is sufficiently large, the CdL instanton disappears and vacuum decay  proceeds via the Hawking-Moss \cite{HM} instanton instead.\footnote{For the possible cosmological signatures of a HM transition, see e.g. \cite{BK}.} In this section we
use the tunneling potential approach to examine this phenomenon.

The decay of a false dS vacuum minimizes the tunneling action (\ref{SVt}) which is the sum of two nonzero pieces, as written in (\ref{SHMCdL}). The first action contribution,  $\Delta S_{HM}$, pulls $V_{0\pp}$ towards $V_\pp$  (and thus $\phi_{0\pp}\to\phi_\pp$) to become smaller, while  $\Delta S_{CdL}$ wants the interval $(\phi_{0\pp},\phi_{0\mm})$ to shrink so as to lower the value of the CdL integral. The final solution comes from the interplay between these two opposing demands \cite{Eg}.  
To get a simple qualitative understanding of this trade-off it is useful to 
make the rough approximation of taking $V_t$ to be a constant.
In that case \cite{Eg}
\be
S(V_t)=\frac{24\pi^2}{\kappa^2}\left(\frac{1}{V_\pp}-\frac{1}{V_t}\right)+\frac{6\pi^2\sqrt{3}}{(\kappa V_t)^{3/2}}\int_{\phi_{0\pp}}^{\phi_{0\mm}}\sqrt{2(V-V_t)}\ d\phi\ ,
\ee
where $\phi_{0\pms}$ are just the two solutions of $V=V_t$.
If one keeps the shape of $V$ fixed but raises the overall energy scale $V_\pp$ then   $\Delta S_{HM}\sim 1/V_\pp^2$ and $\Delta S_{CdL}\sim 1/V_\pp^{3/2}$ and the minimization of the total tunneling action requires $\Delta S_{CdL}\to 0$ with $\phi_{0\pp}\to\phi_{0\mm}\to\phi_B$, the field value for the top of the barrier. In that case, vacuum decay proceeds via the Hawking-Moss instanton with rate 
\be
S=S_{HM}=\frac{24\pi^2}{\kappa^2}\left(\frac{1}{V_\pp}-\frac{1}{V_B}\right)\ ,
\ee
[where $V_B\equiv V(\phi_B)$] as obtained from (\ref{SVt}) when the CdL interval shrinks to zero \cite{Eg}.\footnote{For a derivation of this rate in the context of a stochastic approach to tunneling, see \cite{Vennin}.}

Although the previous behaviour is generic and will happen for potentials of any shape, the particular way in which the transition between CdL-mediated and HM-mediated decay happens depends on the shape of the potential barrier.

\begin{figure}[t!]
\begin{center}
\includegraphics[width=0.45\textwidth]{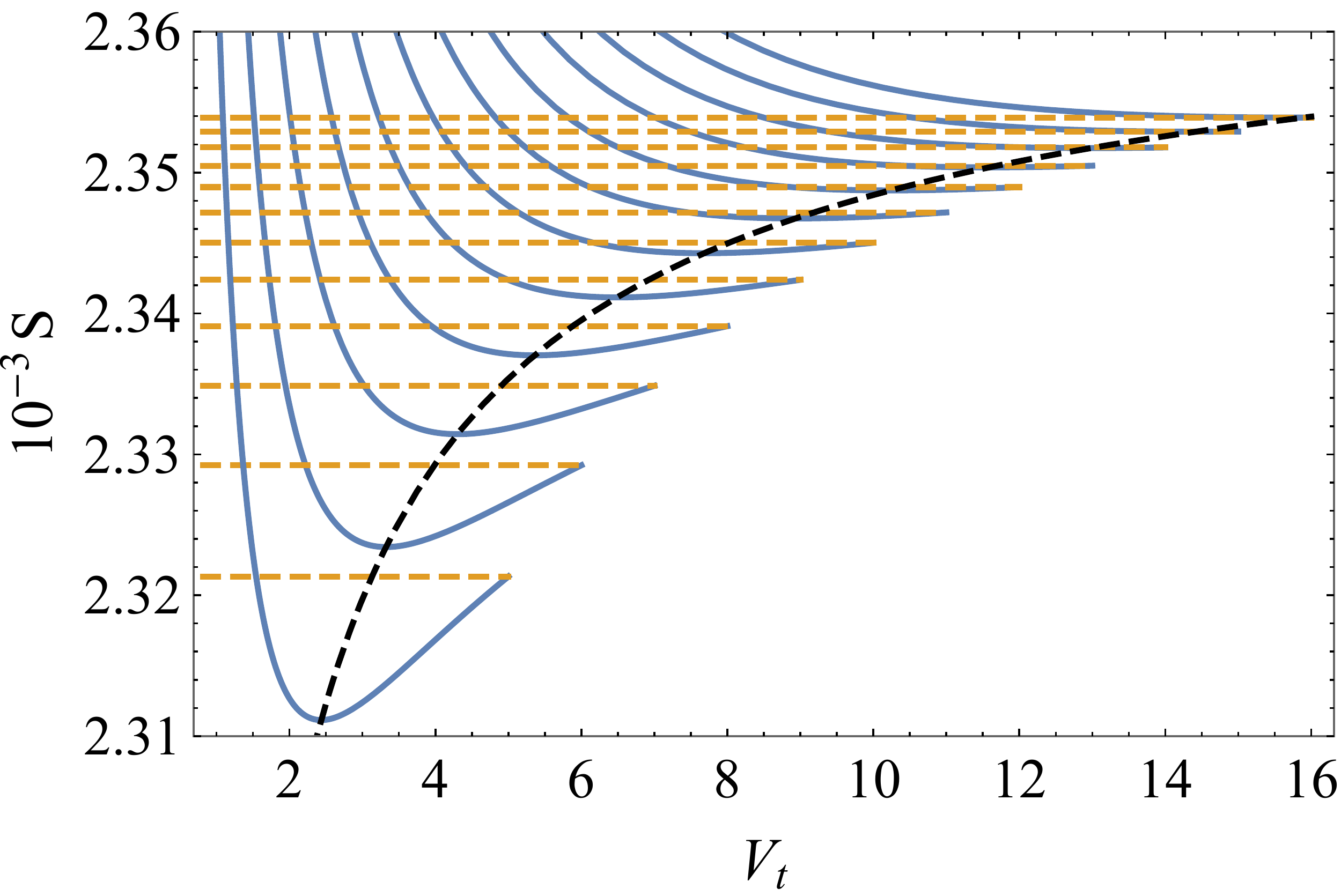}\hspace{0.2cm}
\includegraphics[width=0.45\textwidth]{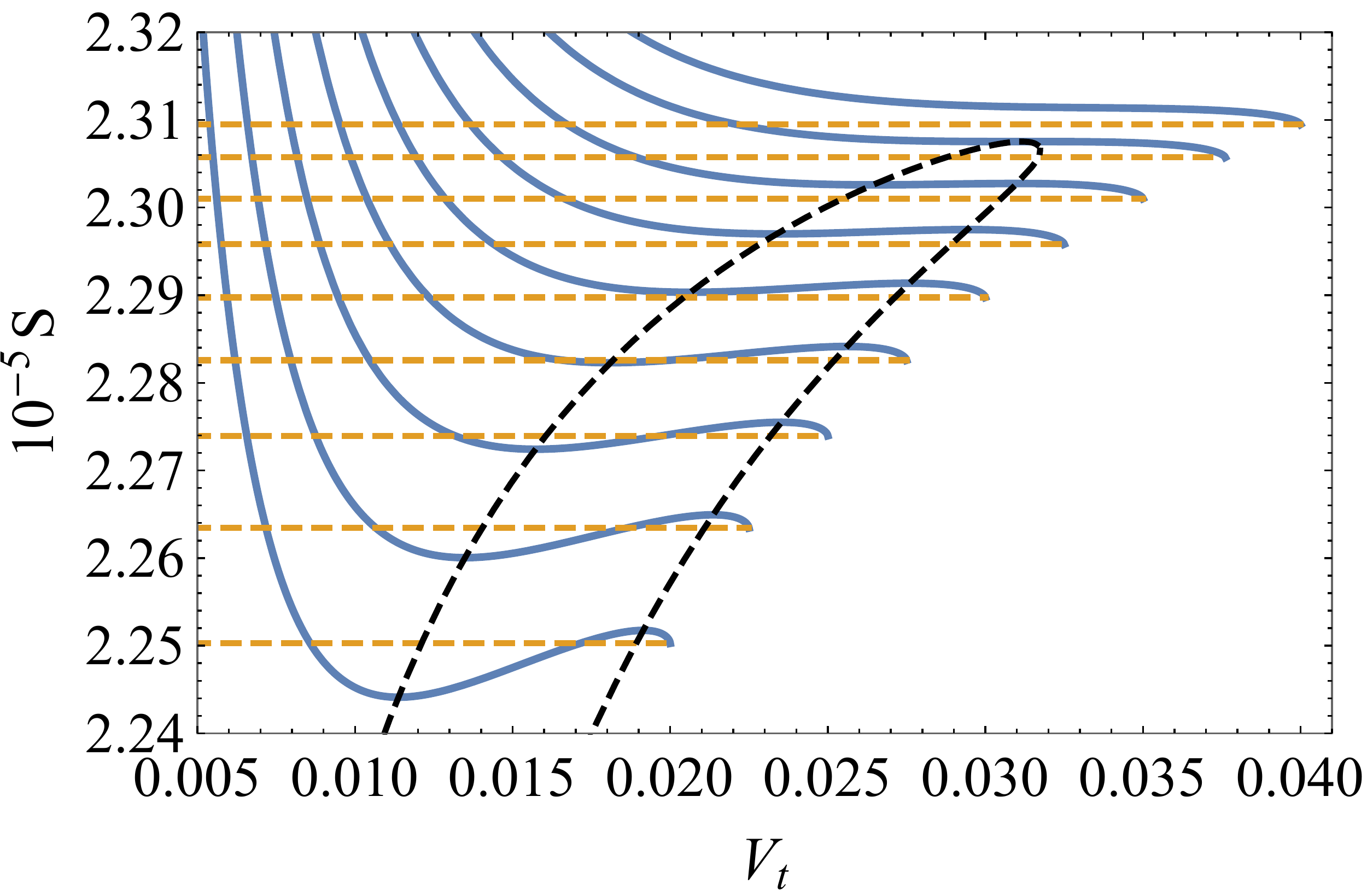}
\end{center}
\caption{Tunneling action $S$ (continuous blue lines) in the constant $V_t$ approximation
for $V=V_B-m^2\phi^2/2$ (left) and $V=V_B-\lambda\phi^4/4$ (right) for different values of $V_B$ ($V_B$ corresponds to the $V_t$ value of the right end of each curve), with the dashed black lines tracking the extrema. The horizontal dashed lines give the corresponding action for HM transitions. We take $\kappa=1$, $V_\pp=0.1\ (0.001)$ for the left (right) plot, $m^2=30$ and $\lambda=1$. 
\label{fig:top}
}
\end{figure}

Consider first the case of a barrier with an inverse quadratic top
\be
V= V_B - \frac12 m^2\phi^2+...
\label{Vquadratic}
\ee
In the constant-$V_t$ approximation $S(V_t)$ can be calculated (and minimized) analytically. One gets
\be
S(V_t)=\frac{24\pi^2}{\kappa^2}\left(\frac{1}{V_\pp}-\frac{1}{V_t}\right)+\frac{6\sqrt{3}\pi^3}{m(\kappa V_t)^{3/2}}(V_B-V_t)\ .
\ee
Figure~\ref{fig:top}, left plot, shows this action (blue continuous curves) as a function of $V_t$ for different values of $V_B$ (the end point of each curve corresponds to $V_t=V_B$). The black dashed line marks the location of the minimum of $S(V_t)$ while the horizontal dashed lines give the value of $S_{HM}$ for each case.
As the scale of the potential $V_B$ increases, the location of the minimum of
$S(V_t)$ gets closer to the top of the barrier $V_B$ and the action tends to the Hawking-Moss one. Above a certain critical value
the minimum of $S(V_t)$ is at $V_B$ and the CdL instanton has disappeared being replaced by the Hawking-Moss one. While we vary $V_B$ we keep $V_\pp$ fixed and this is why the action increases with $V_B$.

The value of the critical $V_B$ calculated in the constant $V_t$ approximation is obtained as $V_{Bc}=16 m^2/(3\kappa \pi^2)\simeq 0.54 m^2/\kappa$ which gives right its order of magnitude. 
The exact value can be obtained easily by solving the EoM  for the exact $V_t(\phi)$ in the limit of CdL disappearance, when $V_t$ gets very close to the top of the barrier.  In that limit, a quadratic expansion should be enough to study the behaviour of $V$ and $V_t$. Writing $V=V_B-m^2\phi^2/2+...$ and $V_t=V_B-\delta -m_t^2\phi^2/2+...$, and imposing the boundary condition $V_t'=3V'/4$ at the $\phi_0\simeq \sqrt{2\delta/(m^2-m_t^2)}$ for which $V_t=V$ one gets 
\be
m_t^2 = \frac34 m^2\ .
\ee 
Using this in the EoM for $V_t$ it follows that $V_{tB}''+\kappa(3V_B-2V_{tB})=0$, from which we get 
the criticality condition
\be
V_{Bc}=\frac{3m^2}{4\kappa}\ .
\label{VBc}
\ee
For $V_B\geq V_{Bc}$ the Hawking-Moss instanton is the relevant one to describe dS vacuum decay. In the analytical examples
we present in section~\ref{sec:ex} it can be checked explicitly how this critical relation is satisfied.
We can also interpret the critical relation (\ref{VBc}) as a limit on how large the curvature at the top should be for the CdL instanton to exist:
\be
m^2> \frac43 \kappa V_B\ .
\label{mbound}
\ee
In the Euclidean formalism, this bound is understood as a necessary condition for the CdL compact bounce to fit inside the horizon at $\phi_B$. Substituting the expressions  for $V$ and $V_t$ above in eq.~(\ref{dphi}) for $\dot\phi$ and integrating that equation
\be
\int_0^{\xi_{CdL}}d\xi = \int_{-\phi_0}^{\phi_0}\frac{d\phi}{\sqrt{2(V-V_t)}}\simeq \frac{1}{\sqrt{m^2-m_t^2}}\int_{-\phi_0}^{\phi_0}\frac{d\phi/\phi_0}{\sqrt{1-\phi^2/\phi_0^2}}\ ,
\ee 
we find that the size of the CdL instanton is $\xi_{CdL}=\pi/\sqrt{m^2-m_t^2}=2\pi/m$. The horizon radius at the top of the barrier, $\phi_B$, is $\xi_{dS,B}=\pi\sqrt{3/(\kappa V_B)}$ and imposing $\xi_{CdL}<\xi_{dS,B}$ gives (\ref{mbound}).

We can then examine what would happen if the top of the potential barrier is flatter, say like an inverted quartic:
\be
V=V_B-\frac14\lambda\phi^4+...
\ee
Again, the constant-$V_t$ approximation can give the rough qualitative behaviour. One gets
\be
S(V_t)=\frac{24\pi^2}{\kappa^2}\left(\frac{1}{V_\pp}-\frac{1}{V_t}\right)+\frac{16\sqrt{3}\pi^2 K(-1)}{\lambda^{1/4}(\kappa V_t)^{3/2}}(V_B-V_t)^{3/4}\ ,
\ee
where $K(m)$ is the complete elliptic integral of the first kind, with $K(-1)\simeq 1.3$.  Figure~\ref{fig:top}, right plot, shows 
$S(V_t)$ for different values of $V_B$ (again keeping $V_\pp$ fixed). For low values of $V_B$, $S(V_t)$ has a minimum with action lower than the HM action (corresponding to the CdL instanton decay) and a maximum with higher action. The location of these minima and maxima is given by the black dashed lines. As $V_B$ increases, the minimum is lifted, until a critical value is reached for which the minimum is degenerate 
with the HM one. For $V_B$ higher than that critical value, the action is minimized by the HM instanton.  For $V_B$ higher than the critical value, the minimum and maximum of $S(V_t)$ get closer until they merge into a saddle point and disappear for even higher $V_B$.

Note that this case is qualitatively different from the previous one with quadratic barrier top for which the CdL-mediated minimum merges with the HM one at the critical value, while now they are still separate. When the barrier top is flat, the CdL integral decreases more slowly as $V_B$ increases and at some point it is advantageous to switch to the HM instanton. The picture obtained above from the simple constant-$V_t$ approximation is confirmed by a numerical calculation of $V_t$.

Potentials with a flat-top barrier have been studied before using the Euclidean approach \cite{JS,BD,HW,BLL},  which emphasized the potential obstruction to the existence of the CdL bounce  when condition (\ref{mbound}) is violated, as well as possible exceptions to that bound.\footnote{Some of these works also discuss the so-called oscillating bounces, which cross the top of the barrier more than once. We do not discuss them here  (although the $V_t$ formalism can be applied to them as well) as they do not seem to be relevant for vacuum decay.}  Although one can also apply the $V_t$ formalism to examine in more detail flat-top potentials, we do not pursue this goal further. We tried to obtain exactly solvable examples of this type of flat-top potentials to include in section~\ref{sec:ex} but we failed to find any.

\section{Exactly Solvable Models: Getting  ${\bma V}$ from ${\bma{V_t}}$ \label{sec:exact}} 
\noindent

The tunneling potential formulation can be quite useful to generate potentials with exactly solvable false vacuum decay.\footnote{Exactly solvable models with gravity included have been obtained using other methods 
before, see \cite{exact0,exact1,exact2} for an incomplete list.
} The idea is to postulate a simple enough $V_t$ and solve its EoM for $V$, which is a much simpler task that solving for $V_t$ given $V$. Without gravity, the integration for $V$ can be done formally, and explicitly for certain $V_t$'s. This was done in \cite{E} to generate some exactly solvable potentials. In the case with gravity the formal integration for $V$ has not been done but \cite{Eg} presented one
exactly solvable case for a particular choice of $V_t$ leading to 
analytic examples of Minkowski or AdS decays. In this section we perform the formal integration for $V$ in the presence of gravity and in the next section we will make use of it to generate a number of examples for dS decays.

Given the expression for $D^2$ in (\ref{D2}) we can formally write $V(\phi)$ in terms of $V_t(\phi)$ and $D(\phi)$
as
\be
V(\phi)=V_t+\frac{D^2-V_t'{}^2}{6\kappa V_t}\ .
\label{VfromD}
\ee
When $D\equiv 0$ the action $S[V_t]$ becomes infinite and the decay is forbidden by gravity. In that case the potential $V$ reproduces the generic form of a critical potential \cite{Estab}. In that sense, $D$ measures deviations of $V$ from criticality.

Using (\ref{VfromD}) the equation of motion (\ref{EoMVt}), given in terms of $V_t$ and $V$, can be rewritten in terms of $D$ and $V_t$ and takes the form
\be
 V_t''+\kappa V_t+\frac{D^2-V_t'{}^2}{2V_t}- V_t'\frac{D'}{D}=0\ .
\label{EoMD}
\ee
This differential equation can be integrated formally to obtain
$D^2$ in terms of $V_t$ as
\be
D^2(\phi)= \frac{V_t'{}^2}{1-V_{t}F}\ ,
\label{D2F}
\ee 
where
\be
F(\phi)\equiv \frac{2\kappa}{E(\phi)} \int_{\phi_0}^{\phi}\frac{E(\tilde\phi)}{V_t'(\tilde\phi)}d\tilde\phi\ ,\quad
E(\phi)\equiv \exp\left[2\kappa\int_{\phi_0}^{\phi}\frac{V_t(\tilde\phi)}{V_t'(\tilde\phi)}d\tilde\phi\right]\ 
\label{FE}
\ee
and $\phi_0$ is some reference field value, taken here to be one of the two contact points between $V$ and $V_t$, so that $D^2(\phi_0)=V_t'{}^2(\phi_0)$. 

The procedure to find an analytical $V$ under control is then to find a $V_t$ simple enough that the integrals $E$ and $F$ can be performed analytically. Once $D^2$ is found, it can be plugged in (\ref{VfromD}) to obtain $V$ as
\be
\boxed{
V(\phi)=V_t+\frac{V_t'{}^2}{6\kappa(1/F-V_t)}}
\label{formalV}
\ee
The function $F$ plays a key role in what follows and it is instructive to rewrite (\ref{EoMD}) as a
differential equation for $F$, which takes the very simple form
\be
F'V_t'=2\kappa(1-FV_t)\ .
\label{EoMF}
\ee
 As is evident from (\ref{VfromD}) and (\ref{D2F}), the zeros of $F$ determine the field values at which $V_t$ touches $V$. In (\ref{formalV}) we see that at such points $1/F$ diverges and the second term drops, giving precisely $V=V_t$. 

The previous results are general: they apply to decays from Minkowski, AdS or dS vacua, and the formalism can be used to obtain analytical potentials in all those cases. In what follows we focus on dS decays.
In order to get sensible results, $V_t$ must satisfy the proper boundary conditions and constraints.  In the case of dS decays, 
the field range of the CdL bounce is $(\phi_{0+},\phi_{0-})$, with  $\phi_{0\pm}$  being different from the minima of the potential. The values of $\phi_{0\pm}$ will correspond to the zeros of $F$:
\be
F(\phi_{0\pm})=0\ .
\label{phi0fix}
\ee
This condition is crucial as it fixes the value of $\phi_{0\pm}$. We show explicit examples of this in the next section.

In the dS case, $V_t$ is not monotonic but rather it should have a maximum at some intermediate field value $\phi_{T}$ (which corresponds to the maximum value of the metric function $\rho$ in the Euclidean formulation, see \cite{Eg})  and at this maximum the top of $V_t$ should be an inverted quadratic:
 From the EoM (\ref{EoMVt}) it follows that 
 \be
 V_{tT}''=-\kappa(3V_T-2V_{tT})<0\ .
 \ee 
 (The subindex $T$ on a  function indicates that it is evaluated at $\phi_T$.)
 Eq.~(\ref{formalV}) shows that  $V_t'=0$ would naively lead to $V_T\equiv V(\phi_T)=V_{tT}\equiv V_t(\phi_T)$ while the contact points between $V$ and $V_t$ should be just $\phi_{0\pm}$. The condition $V_T\neq V_{tT}$ requires 
\be
F(\phi_T)V_t(\phi_T)=1\ ,
\label{FVt1}
\ee
so that the ratio in (\ref{formalV}) gives a finite value. This means 
$1-FV_t$ goes to zero at $\phi_T$ as $V_t'{}^2$ does. From (\ref{EoMF}) we then learn that $F'$ also vanishes at $\phi_T$.
Using (\ref{EoMF}), (\ref{FVt1}) and L'H\^opital's rule we get
\be
V_T=V_{tT}+\frac{V_{tT}''}{3V_{tT}F_T''}\ .
\label{VTexp}
\ee
Moreover, (\ref{D2F}) shows that $FV_t<1$ and therefore (\ref{EoMF}) forces $F'$ to have the same sign as $V_t'$: $F$ is also a nonmonotonic function with a maximum and $V_{tT}''/F_T''>0$, with $F_T''<0$ and $V_{tT}''<0$.

In the next section we put this formalism to work and obtain several examples of analytical potentials describing dS to dS transitions  (as well as one transition from dS to AdS).  We relegate to the  appendix a simple case of Minkowski false vacuum decay, that generalizes the scale invariant potential $V(\phi)=-\lambda\phi^4/4$ to the case with gravity.

\section{Examples of Solvable  dS Decays\label{sec:ex}}

To use the formalism described in the previous section we can postulate a simple $V_t$ (with a maximum at some point $\phi_T$) that allows to integrate explicitly (\ref{EoMF}) for $F$. The result will depend on an integration constant $C$ that is fixed in the following way.  A Taylor expansion of $F$ around $\phi_T$ takes the generic form
\be
F(\phi_T+\epsilon) = \frac{1}{V_{tT}}+\frac{g^2}{\kappa-\kappa_c}\epsilon^2+{\cal O}(\epsilon^3)+(C-C_\kappa){\cal O}(\epsilon^{2\kappa/\kappa_c})\ ,
\label{Fexp}
\ee
with $g^2,\kappa_c>0$,  $C_\kappa$ is some model dependent constant and
\be
\kappa_c=-V_{tT}''/V_{tT}\ .
\ee
As explained in the discussion at the end of the previous section, around eq.~(\ref{VTexp}), we need $F''(\phi_T)<0$ and finite to have $V_t$ below $V$ at $\phi_T$. From (\ref{Fexp}) we immediately see that to satisfy that requirement we need to respect the limit $\kappa<\kappa_c$ (if this is violated the decay occurs via the HM instanton rather than the CdL bounce) and fix the integration constant as $C=C_\kappa$.

Although most of the examples we present below follow this method, there is an alternative way to generate solvable potentials:  postulate a simple enough $F$, with two zeros and a maximum between them, and integrate equation (\ref{EoMF}) for $V_t$. Example~\ref{7.4} illustrates this route.  Besides the examples we present below we have found many others and the interested reader should be able to produce new ones easily.

\subsection{$\bma{V_t(\phi) =A-1+ \cos\phi}$\label{7.1}}
For simplicity, in this $V_t$ (and some other examples below) we suppress mass scales (e.g. in the normalization of the field and the prefactor of the cosine, etc.). These scales can be recovered trivially if needed. As $V_t$ is symmetric under $\phi\to-\phi$, the potential $V$ enjoys the same symmetry. The CdL field range will simply be $(\phi_{0+},\phi_{0-})=(-\phi_0,\phi_0)$ with $0<\phi_0<\pi$ and the maximum of $V_t$ occurs at $\phi_T=0$. 

Equation~(\ref{EoMF}) can be integrated analytically and $F$ expressed in terms of a hypergeometric function as 
\be
F(\phi)=\frac{[\cos(\phi/2)]^{4\kappa}}{A}\, _2F_1[-2\kappa,-A\kappa;1-A\kappa;-\tan^2(\phi/2)]+C \frac{(\sin^2\phi)^\kappa}{[\tan^2(\phi/2)]^{(1-A)\kappa}}\ ,
\ee
where $C$ is an integration constant. Expanding $F(\phi)$ around $\phi_T=0$ we have
\be
F(\epsilon)=\frac{1}{A} + \frac{\kappa}{2A(A\kappa-1)}\epsilon^2+{\cal O}(\epsilon^3)+
 C\, \epsilon^{2A\kappa}\left[4^{(1-A)\kappa}+{\cal O}(\epsilon^2)\right]\ ,
\ee 
which conforms to the generic expression (\ref{Fexp}). We immediately conclude that $C=0$ and the condition to have a CdL bounce is $\kappa<\kappa_c=1/A$. If that bound is not satisfied the transition occurs via the Hawking-Moss instanton.
Having fixed $F$, the potential is then obtained by plugging $F$ above on (\ref{formalV})
and $\phi_0$ is obtained from $F(\phi_0)=0$.

\begin{figure}[t!]
\begin{center}
\includegraphics[width=0.45\textwidth]{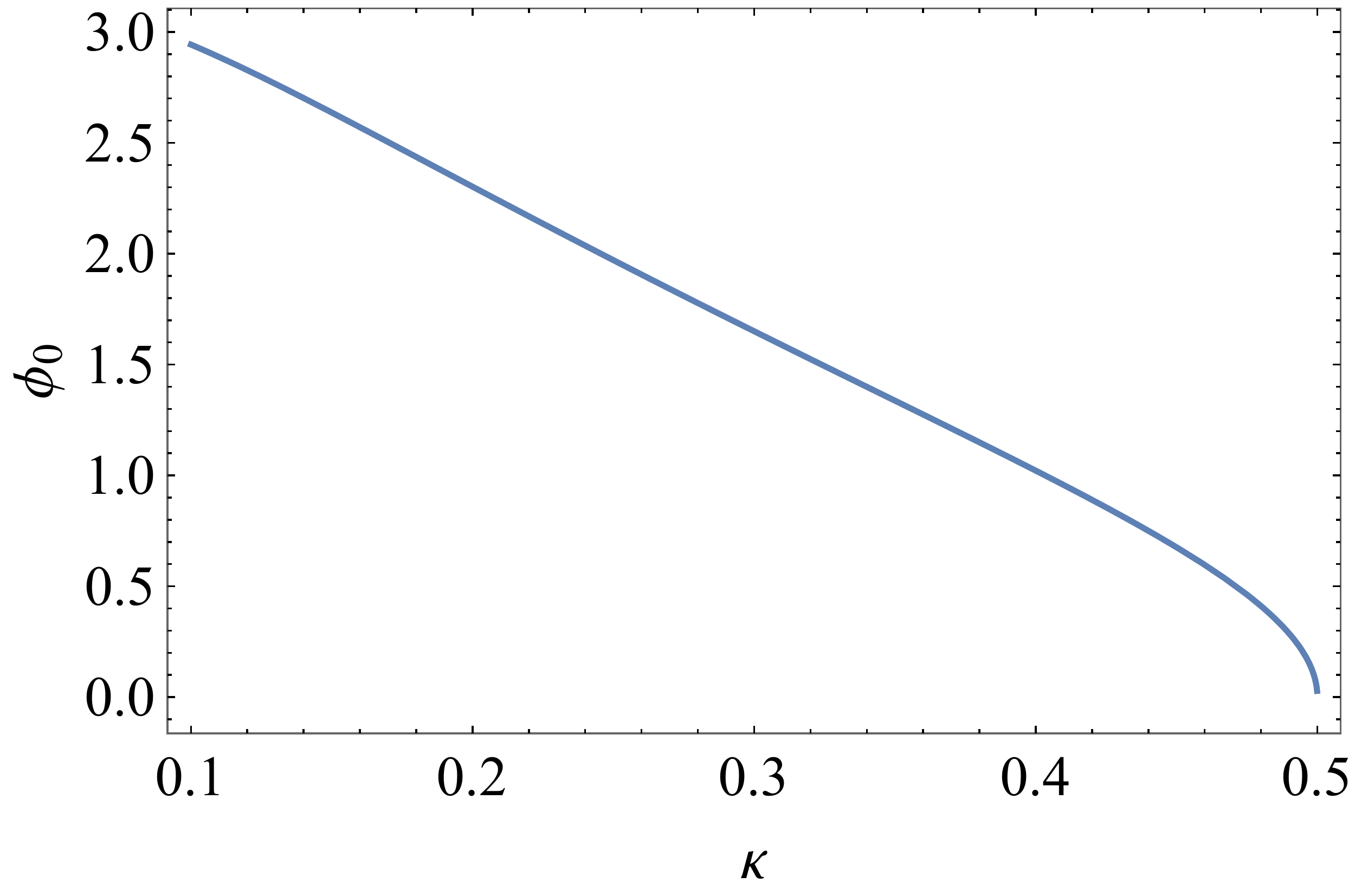}\hspace{0.2cm}
\includegraphics[width=0.45\textwidth]{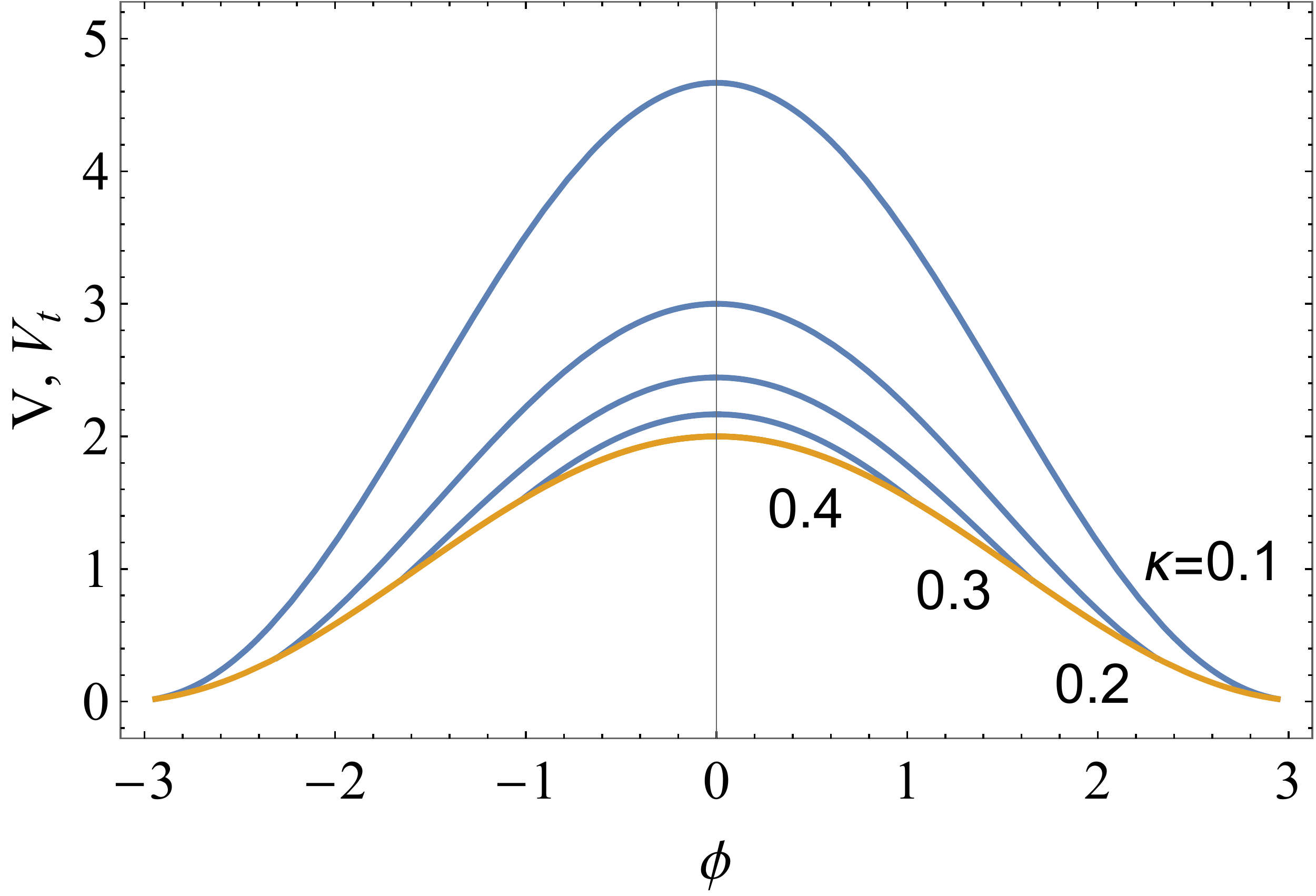}
\end{center}
\vspace{-0.5cm}
\caption{
For example \ref{7.1} with $A=2$, varying $\kappa$. Left plot: Limit, $\phi_0$, of the CdL interval. Right plot: For the fixed $V_t$,  several $V$'s in the CdL range for the indicated values of $\kappa$.
\label{fig:71b}
}
\end{figure}

For numerics, let us examine first the case with fixed $A$ (we take $A=2$) and let $\kappa$ vary. Figure~\ref{fig:71b} shows $\phi_0$ as a function of $\kappa$ and different $V$'s for several choices of $\kappa$, as indicated. 
The numerical analysis confirms  the critical value $\kappa_{c}=1/A=1/2$, at which the CdL range shrinks to zero ($\phi_0\to 0$). Above that critical value the transition occurs via the Hawking-Moss instanton.

\begin{figure}[t!]
\begin{center}
\includegraphics[width=0.45\textwidth]{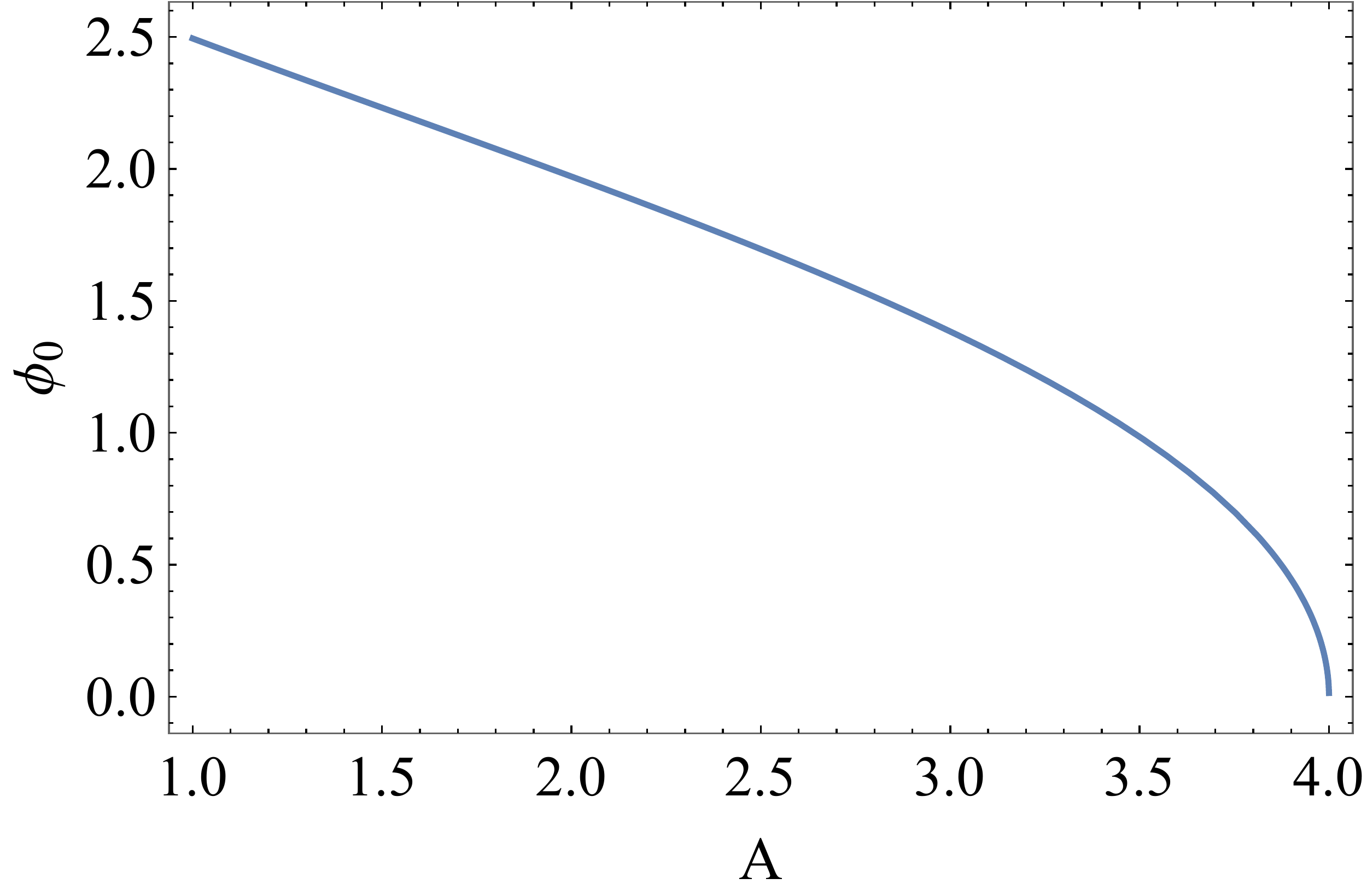}\hspace{0.2cm}
\includegraphics[width=0.45\textwidth]{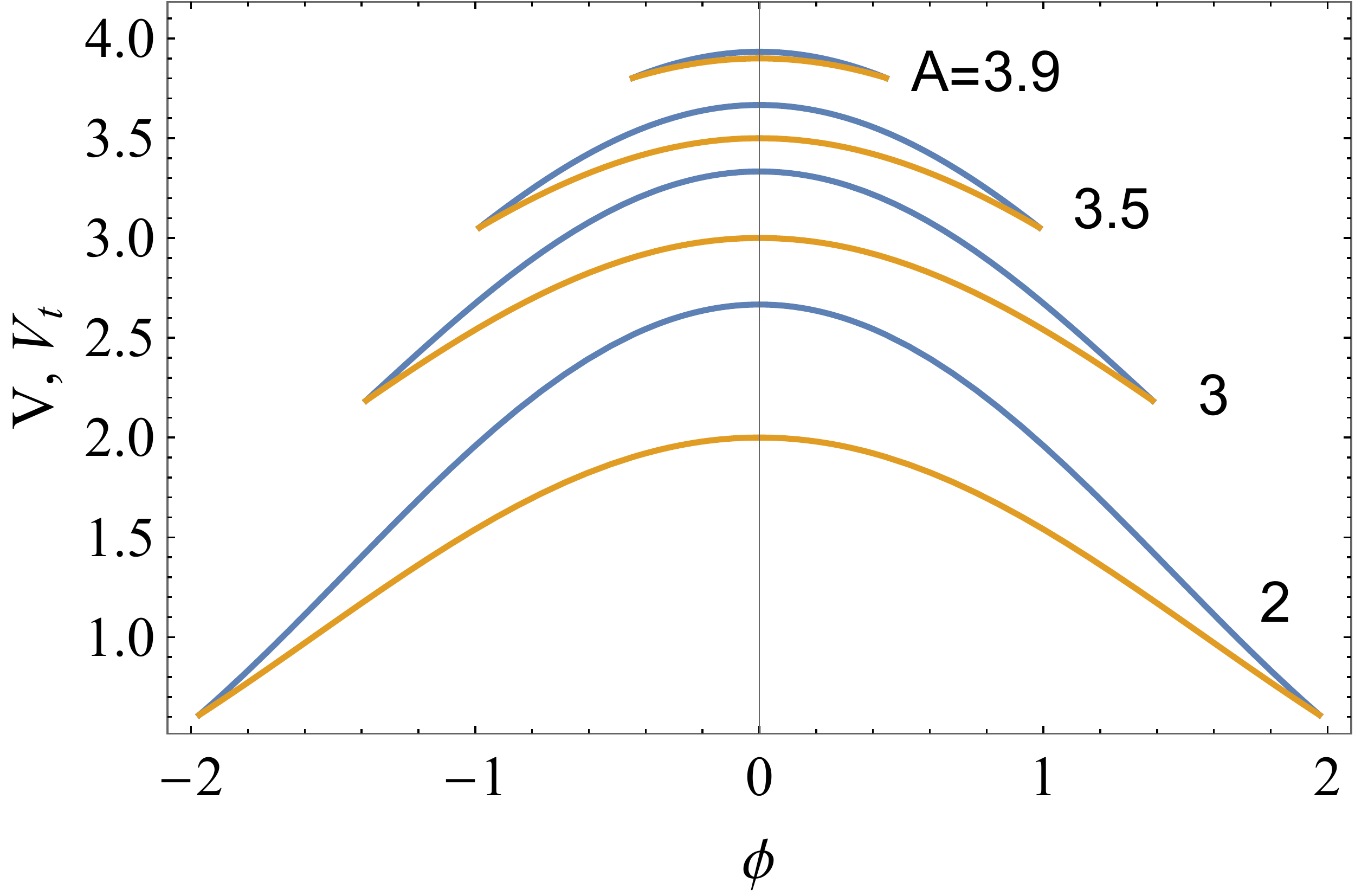}
\vspace{-0.5cm}
\end{center}
\caption{For example \ref{7.1} with $\kappa=1/4$, varying $A$. Left plot: Limit, $\phi_0$, of the CdL interval. Right plot: $V$ and $V_t$ pairs in the CdL range for the indicated values of $A$.
\label{fig:71a}
}
\end{figure}

\begin{figure}[t!]
\begin{center}
\includegraphics[width=0.485\textwidth]{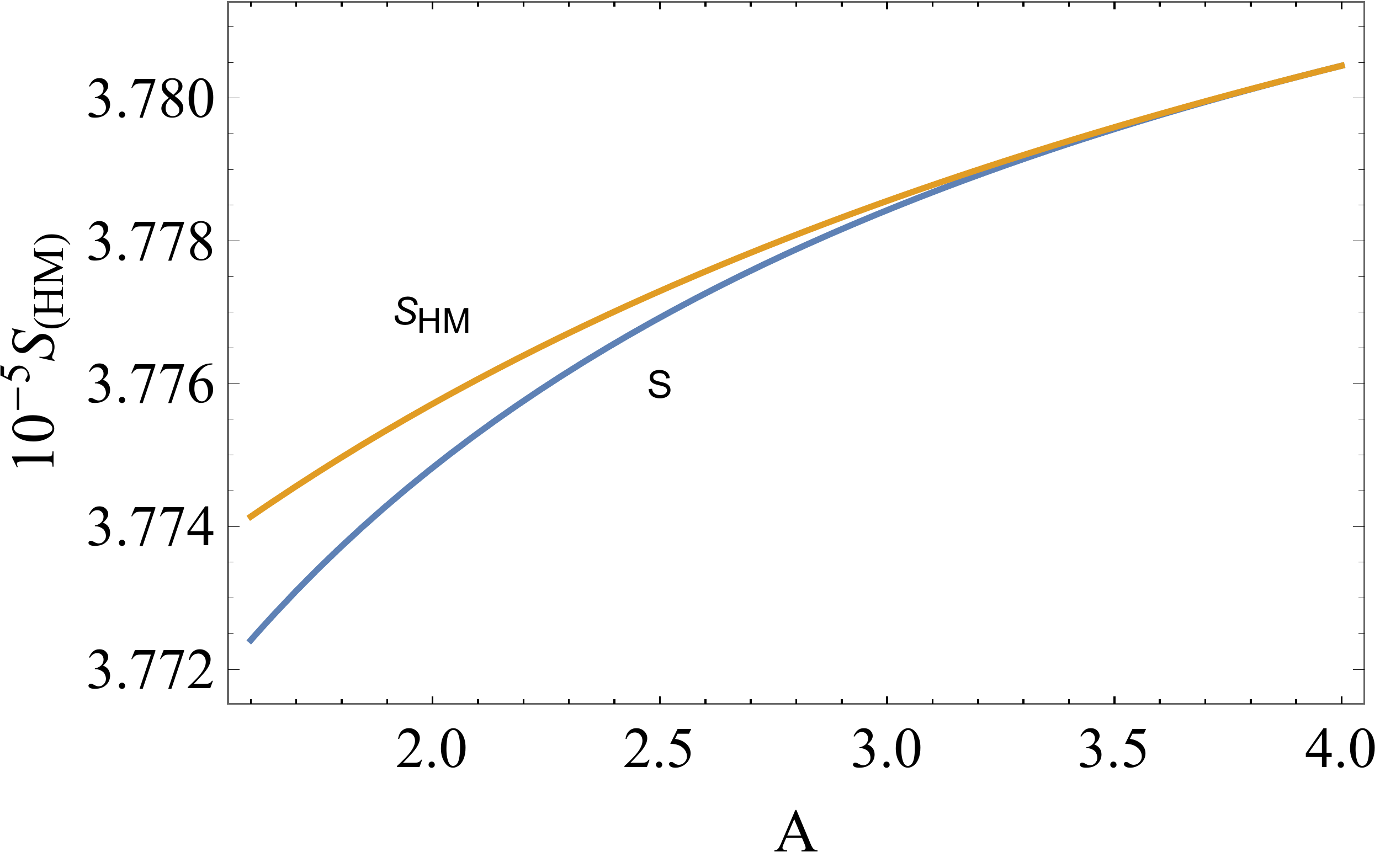}
\includegraphics[width=0.45\textwidth]{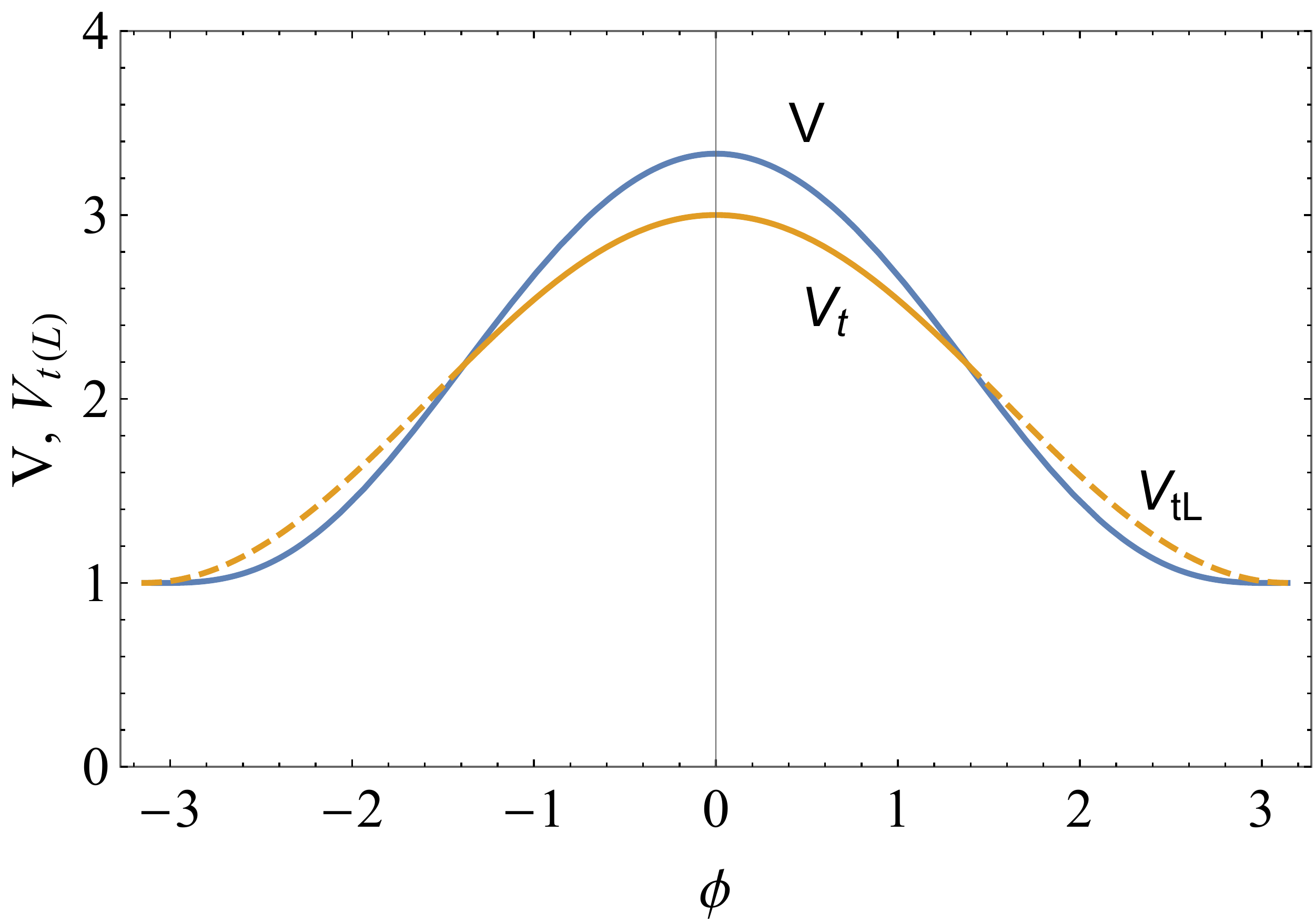}
\end{center}
\vspace{-0.5cm}
\caption{For example \ref{7.1}, right plot: Actions for decay via HM transition or via CdL bounce as a function of $A$, with $\kappa=1/4$. Left plot: Potential and tunneling potential, extended beyond the CdL range,  with $A=3$ and $\kappa=1/4$. The extension of $V_t$ is labeled $V_{tL}$ and shown by dashed lines. 
\label{fig:VtL}
}
\end{figure}

It is also instructive to fix $\kappa=1/4$ and let $A$ vary. 
Figure~\ref{fig:71a} shows $\phi_0$ as a function of $A$ and pairs of $V$ and $V_t$ for several choices of $A$, as indicated. We see that, as expected, there is a critical value for $A$, $A_{c}=1/\kappa=4$, at which the CdL range shrinks to zero ($\phi_0\to 0$), in accordance with the bound $\kappa<1/A$  derived above. 
In figure~\ref{fig:VtL} we show, for the same choice of parameters, 
the actions for decay via the HM transition or via the CdL bounce
(fixing $V_\pp=0.01$ for concreteness), confirming that the latter action is smaller. This is not guaranteed (see discussion in section~\ref{sec:HM}) and must be checked in each case. The rest of the examples we present do pass this test.


When a potential $V$ is obtained from a postulated $V_t$, as done in this section, in principle $V$ is defined only in the CdL interval $({\phi_{0\pp},\phi_{0\mm}})$. Outside that interval one is free to
extend $V$ almost arbitrarily, although the continuity of $V$ and $V'$ at $\phi_{0\pm}$ are natural conditions to impose.
In addition, the location of the potential minima should be compatible with the decay considered. For instance, a $V_t$ with a maximum describes dS decay and therefore, the false vacuum
(in the region of $V$ outside the CdL interval) should have $V_\pp>0$.  

In some cases, the function $V_t$ that has been used to construct $V$ is defined and well behaved outside the CdL interval and can be used to extend also $V$ in that field range. Figure~\ref{fig:VtL} illustrates this, for $A=3$ and $\kappa=1/4$, with $V$ and $V_t$ both simply extended beyond the CdL range. The extension of $V_t$, labeled $V_{tL}$ in the figure, can be simply ignored or if not, needs to be reinterpreted, as it breaks the crucial relation $V_t=V-\dot\phi^2/2$ which implies $V_t\leq V$. This extension is
similar to the Lorentzian continuation of Euclidean CdL geometries performed in \cite{exact0} to extend the exact solutions to describe the aftermath of bubble nucleation. 
Exploring the precise meaning of this extension for $V_t$ would be interesting but goes beyond the goals of the present paper.

\subsection{$\bma{V_t(\phi)=e^{-\phi^2}}$\label{7.2}}
For this $V_t$, integrating (\ref{EoMF}),
we get
\be
F(\phi)\equiv \ \frac{\kappa}{2}(-\phi^2)^{\kappa/2}\left[\Gamma(-\kappa/2,-\phi^2)-\Gamma(-\kappa/2)\right] +C (\phi^2)^{\kappa/2} \ ,
\ee
where $\Gamma(a,z)$ is the incomplete gamma function. Expanding $F(\phi)$ around $\phi_T=0$ one gets
\be
F(\epsilon)=2+\frac{\kappa}{\kappa-2}\epsilon^2+{\cal O}(\epsilon^3)+ C (\epsilon^2)^{\kappa/2}\ ,
\ee
from which we read $\kappa_c=2$, $C=0$, in accordance with the general discussion around (\ref{Fexp}).
The potential is then obtained by plugging $F$ above on (\ref{formalV}) and $\phi_0$ is obtained from $F(\phi_0)=0$.

\begin{figure}[t!]
\begin{center}
\includegraphics[width=0.4\textwidth]{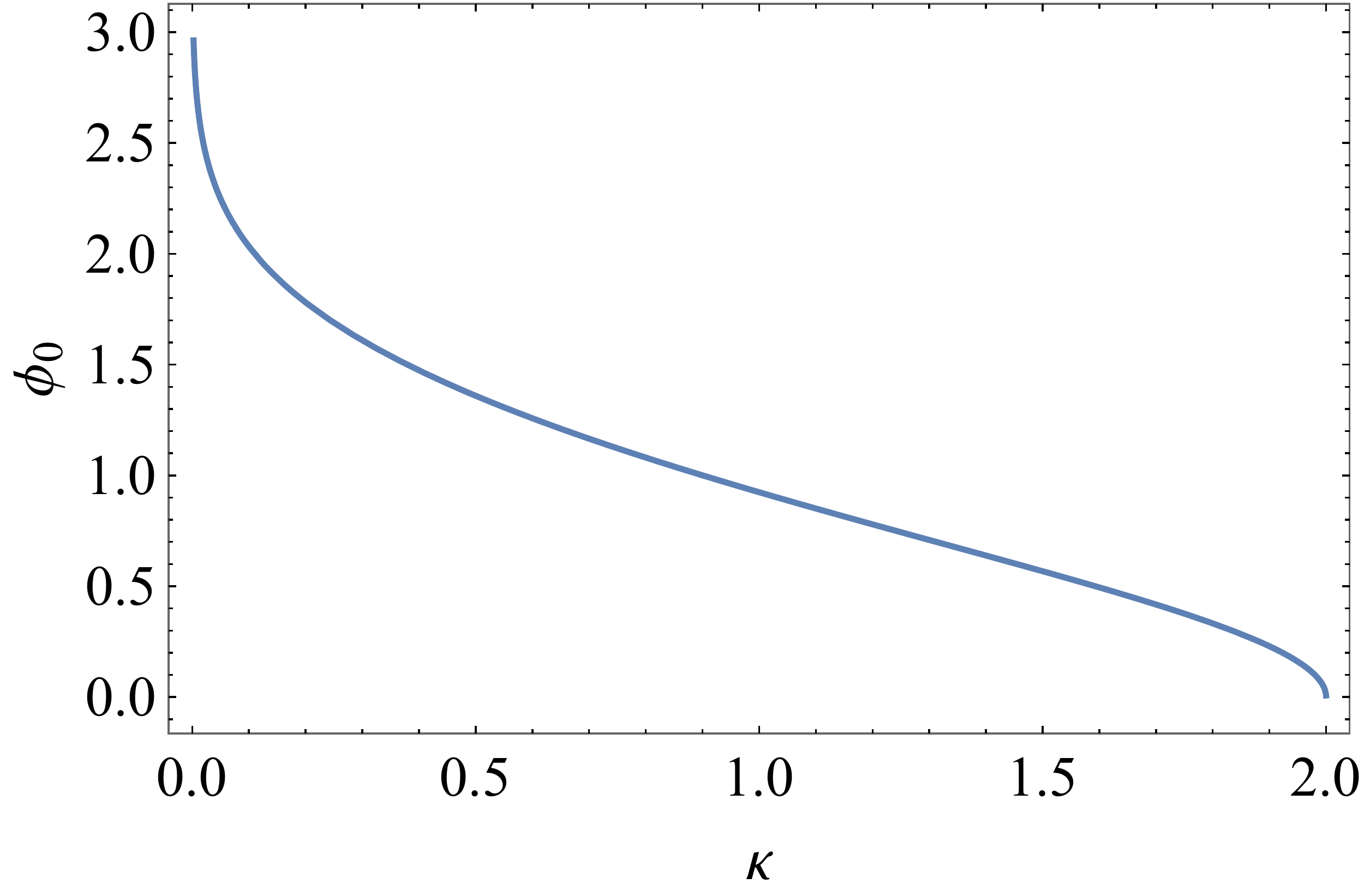}\hspace{0.2cm}
\includegraphics[width=0.4\textwidth]{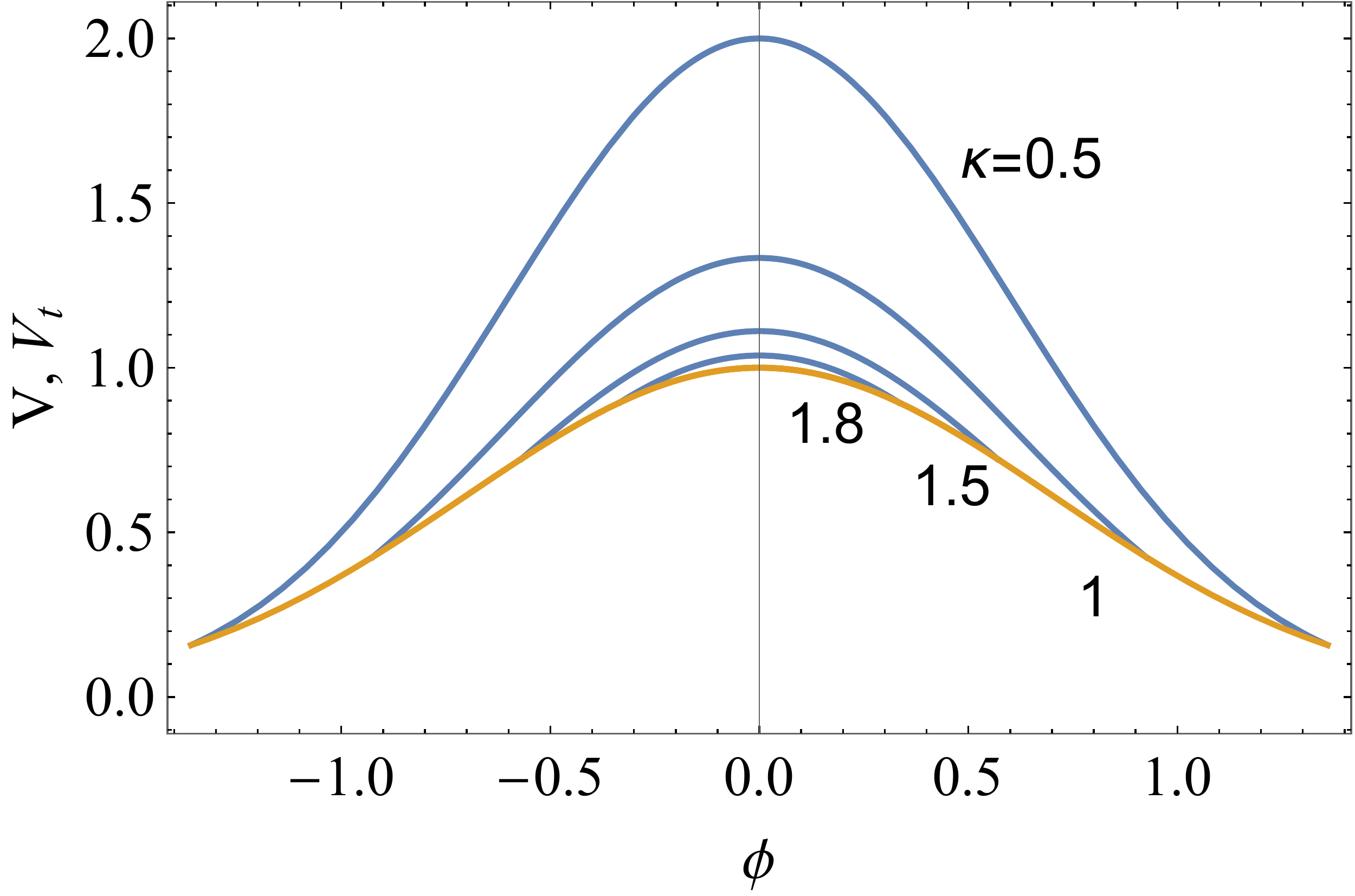}
\vspace{-0.5cm}
\end{center}
\caption{For example \ref{7.2}, varying $\kappa$. Left plot: Limit, $\phi_0$, of the CdL interval. Right plot: For the fixed $V_t$,  several $V$'s in the CdL range for the indicated values of $\kappa$.
\label{fig:72}
}
\end{figure}

Figure~\ref{fig:72} shows $\phi_0$ as a function of $\kappa$ and different $V$'s for the same $V_t$ for several choices of $\kappa$, as indicated. We see how the CdL range shrinks to zero ($\phi_0\to 0$) as $\kappa\to\kappa_c$. Above that critical value the transition occurs via the Hawking-Moss instanton.

\begin{figure}[t!]
\begin{center}
\includegraphics[width=0.34\textwidth]{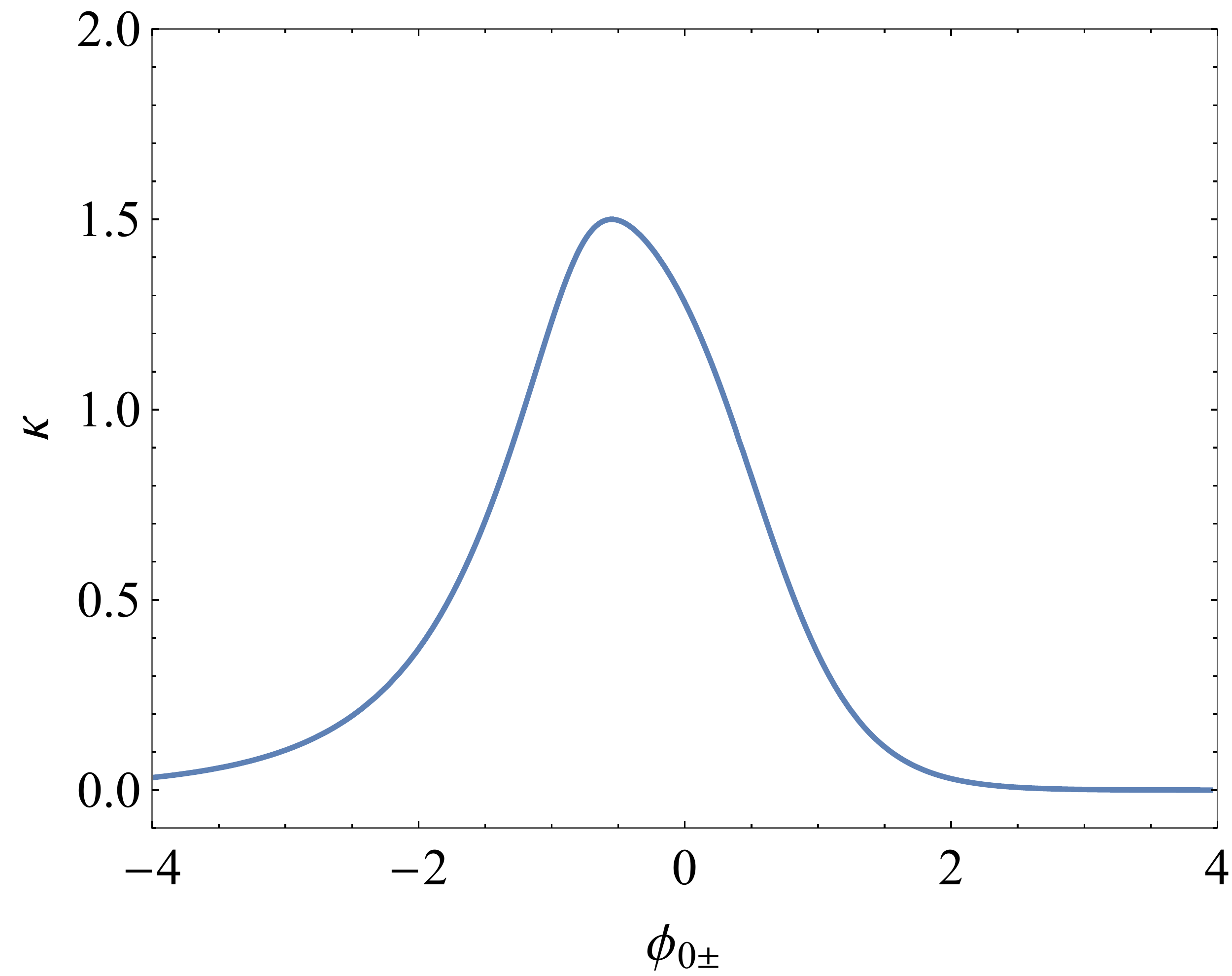}\hspace{0.2cm}
\includegraphics[width=0.4\textwidth]{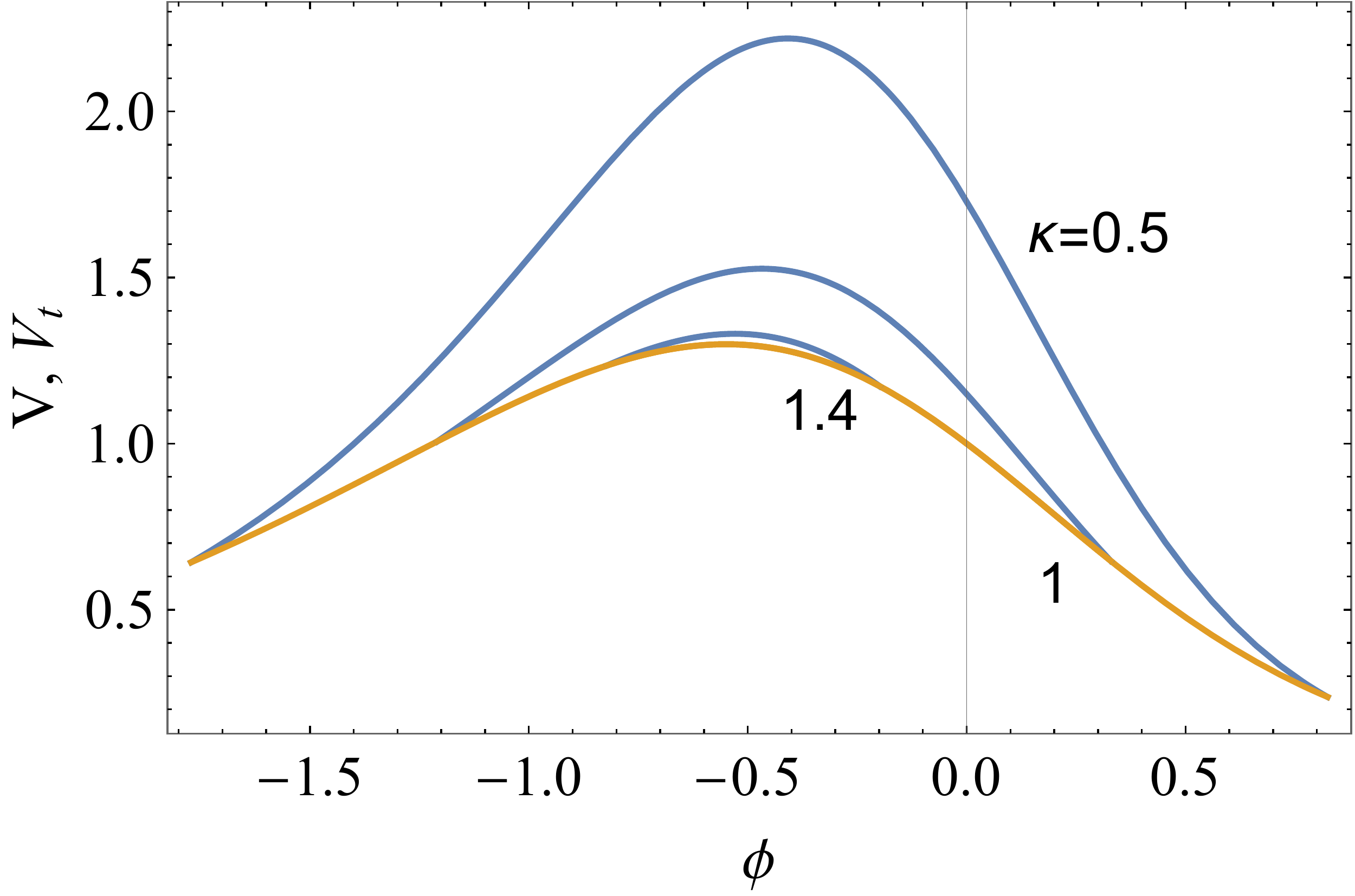}
\vspace{-0.5cm}
\end{center}
\caption{For example \ref{7.3}, varying $\kappa$. Left plot: Limits, $\phi_{0\pm}$, of the CdL interval. Right plot: For the fixed $V_t$,  several $V$'s in the CdL range for the indicated values of $\kappa$.
\label{fig:73}
}
\end{figure}

\subsection{$\bma{V_t(\phi)=e^{-\phi}/\cosh^2\phi}$\label{7.3}}
This is one simple example  with $V_t$ and $V$ that are not symmetric under $\phi-\phi_T\to\phi_T-\phi$.
For these cases the critical equation $F(\phi)=0$ has two non symmetric zeros,  $\phi_{0\pm}$, that give the extremes of the CdL interval. For the example above we have\footnote{For the particular case $\kappa=3/4$, $F(\phi)$ can be expressed in terms of elementary functions, although it is not particularly simple.}
\bea
F(\phi)&=&\frac{\kappa}{\sqrt{3}(9-2\kappa)(9-4\kappa^2)}\left\{\frac{1}{6\sqrt{z}}\left[9(111-30z-z^2)-144\kappa(1+z)+4\kappa^2(3+z)^2\right]
\right.\nonumber\\
&+&\left.
48(1-z)^{4\kappa/3}z^{-\kappa}(3-4\kappa)
\left[B(z;\kappa-1/2,-4\kappa/3)-C\right]\right\}\ ,
\eea
where $z\equiv 3\,e^{2\phi}$ and $B(z;a,b)$ is the incomplete beta function. Expanding $F(\phi)$ around
the maximum of $V_t$, at $\phi_T=-\log(3)/2$, we get
\be
F(\phi_T+\epsilon)=\frac{4}{3\sqrt{3}}+\frac{2\kappa\epsilon^2}{\sqrt{3}(2\kappa-3)}+{\cal O}(\epsilon^3)+{\cal O}(\epsilon^{4\kappa/3})
\left[C-C_\kappa+{\cal O}(\epsilon)\right]\ ,
\ee
with
\be
C_\kappa=-\frac{\pi\csc(4\pi\kappa/3)\Gamma(\kappa-1/2)}{\Gamma(-1/2-\kappa/3)\Gamma(1+4\kappa/3)}\ .
\ee 
From this expansion we obtain $\kappa_c=3/2$ and $C=C_\kappa$.
 
The potential is obtained from (\ref{formalV}) and the extremes of the CdL interval come from solving $F(\phi_{0\pm})=0$. They
are shown in figure~\ref{fig:73}, left plot, with $\phi_{0+}=\phi_{0-}$ for $\kappa=3/2$. The right plot gives $V_t$ and $V$ for several values of $\kappa$ as indicated, showing again how for $\kappa\to\kappa_c=3/2$
the CdL interval disappears.

\subsection{$\bma{V_t(\phi)=2\left(\cos\theta+\cos\phi\right)/\sin^2\theta}$\label{7.4}}
As mentioned, we can also get simple examples by postulating a simple $F$, from which one gets $V_t$ and then $V$. For instance, if we take $F=a V_t+b$ and solve for $V_t$ we end up with the simplest example we have found. We get
\be
V_t(\phi)=\frac{1}{4a}\left[-2b+(1+4a+b^2)\cos(\sqrt{2\kappa}\phi+C)+i (1-4a-b^2)\sin(\sqrt{2\kappa}\phi+C)\right]\ ,
\ee
where $C$ is an integration constant.
Setting $\kappa=1/2$,  $C=0$ and $b=-\sqrt{1-4a}$ (to have a real $V_t$), we end up with 
\be
V_t(\phi)=\frac{2}{\sin^2\theta}\left(\cos\theta+\cos\phi\right)\ ,
\ee
where we have introduced the angular parameter $\theta$, given by $4a=\sin^2\theta$, with $0\leq \theta\leq \pi/2$. From this we obtain the axion-like potential
\be
V(\phi)=\frac{4}{3\sin^2\theta}\left(\cos\theta+2\cos\phi\right)\ ,
\ee
and the CdL range is simply the interval $(-\theta,\theta)$, with $V(\pm\theta)=V_t(\pm\theta)=4\cos\theta/\sin^2\theta$ at the contact points of $V$ and $V_t$.
The CdL part of the action can be calculated exactly as
\be
\Delta S_{CdL}=24\pi^2\sin\theta\,(\tan\theta-\theta)\ .
\ee
The limit $\theta\to 0$ corresponds to the CdL instanton disappearance (which in this particular example only happens for $V_B=V(0)\to\infty$) with $\Delta S_{CdL}\to 0$. The limit $\theta\to\pi/2$ corresponds instead to a forbidden Minkowski to Minkowski  transition and has $\Delta S_{CdL}\to\infty$.

\begin{figure}[t!]
\begin{center}
\includegraphics[width=0.4\textwidth]{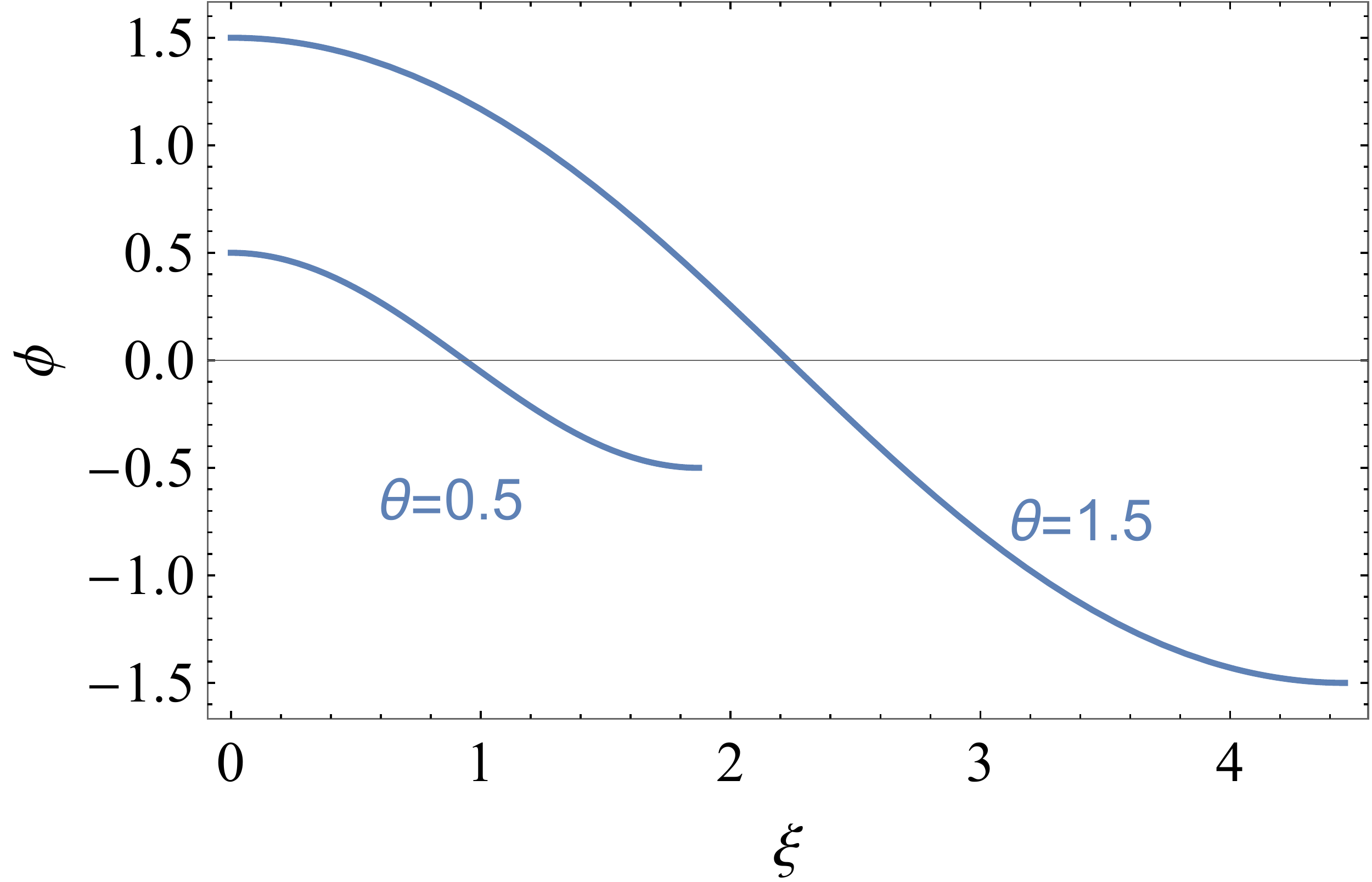}\hspace{0.2cm}
\includegraphics[width=0.4\textwidth]{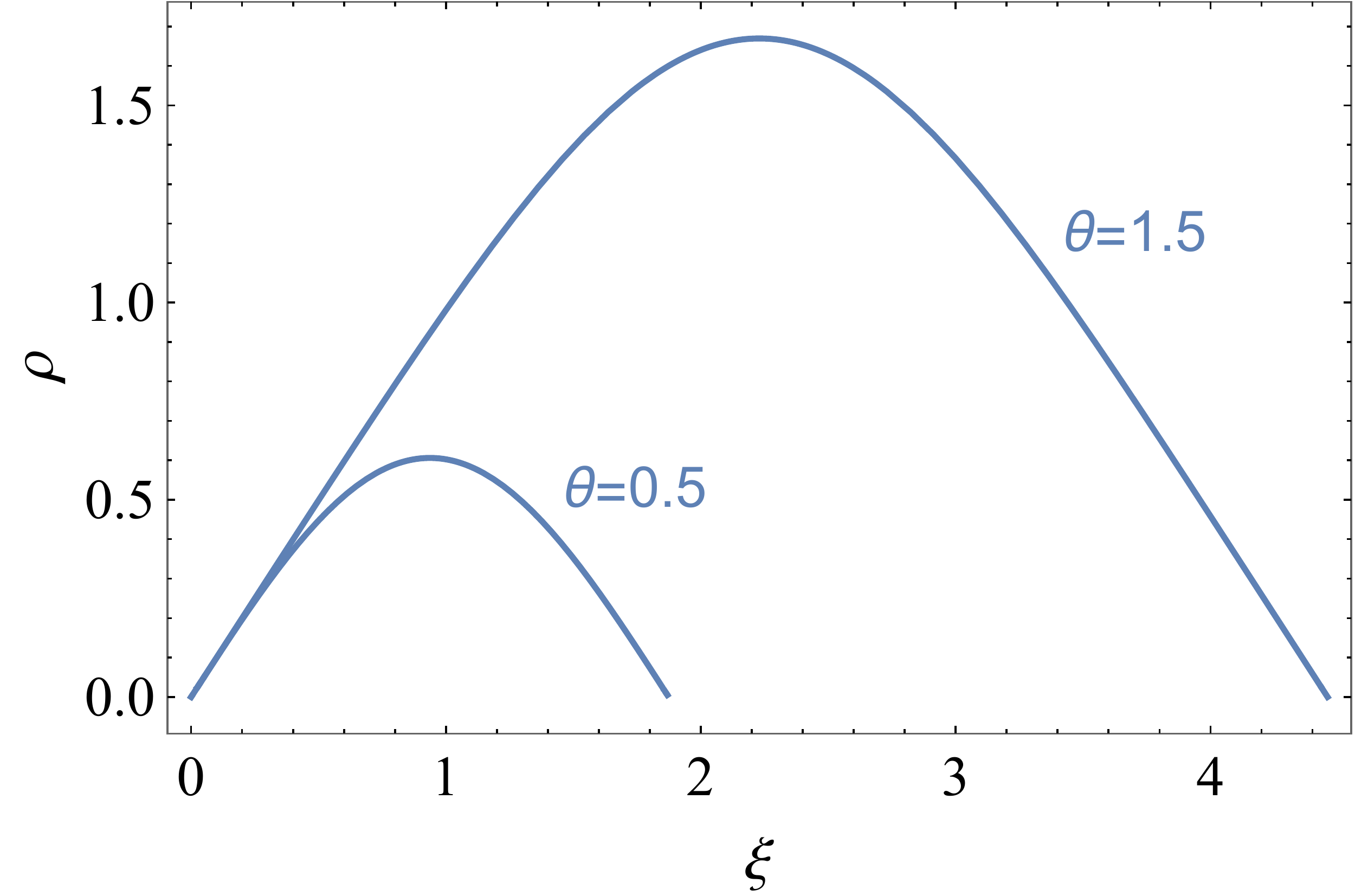}
\end{center}
\caption{For example~\ref{7.4} and the indicated values of $\theta$. Left plot: Field profile $\phi(\xi)$ of the Euclidean bounce. Right plot: Profile of the Euclidean metric function $\rho(\xi)$.
\label{fig:phirho}
}
\end{figure}

It is instructive to derive the field and metric Euclidean profiles, $\phi(\xi)$ and $\rho(\xi)$ for this simple example. By integrating $d\phi/d\xi=-\sqrt{2(V-V_t)}$, see (\ref{dphi}), we get
\be
\phi(\xi)=-2\ \mathrm{am}\left(\left.\frac{2\xi-\xi_e}{2\sqrt{6}\cos(\theta/2)}\right|\csc^2(\theta/2)\right)
\ee
where $\mathrm{ am}(u|m)$ is the Jacobi amplitude function
and 
\be
\xi_e=\sqrt{6}\sin\theta\ K(\sin^2(\theta/2))\ ,
\ee
with $K(m)$ the complete elliptic function of the first kind.  The metric function is obtained from $\rho=3\sqrt{2(V-V_t)}/D$, see (\ref{rho}), as
\be
\rho(\xi)=\sqrt{3[\cos\phi(\xi)-\cos\theta]}\ .
\ee
The finite interval over which the CdL instanton is defined is $\xi\in (0,\xi_e)$, with $\rho$ being zero at both extremes. Figure~\ref{fig:phirho} shows the profiles of both $\phi(\xi)$ and $\rho(\xi)$ for two choices of the $\theta$ parameter. This example nicely illustrates the complementarity between the Euclidean and $V_t$ formalisms showing how a complicated description in one case can become elementary in the other.

\subsection{$\bma{V_t(\phi)=A e^{\phi}-e^{a\phi}}$\label{7.5}}

This example, with $A>0$ and $a>1$, is interesting as
it can feature a dS vacuum decaying to AdS.
This $V_t$ peaks at
 $\phi_{T}=\log(A/a)/(a-1)$ and is
an asymmetric example with two different zeros $\phi_{0\pm}$.
The function $F$ is obtained as
\be
F(\phi)=\frac{r}{q(a-1)}\left[\left(
\frac{a}{A}\right)^q\left(e^{-\phi}-\frac{A}{a}e^{-a\phi}\right)^rC+e^{-a\phi} {}_2F_1\left(1,p;1+q;\frac{A}{a}e^{-(a-1)\phi}\right)
\right]\ ,
\ee
where $r=2\kappa/a$, $p=(a-2\kappa)/(a-1)$, $q=1+p/a$ while ${}_2F_1(a,b;c;z)$ is the hypergeometric function and $C$ is an integration constant. 
The expansion of $F$ around $\phi_T$ reads
\be
F(\phi_T+\epsilon)=\frac{1}{a-1}\left(\frac{a}{A}\right)^{a/(a-1)}
\left[1+\frac{a\kappa \epsilon^2}{2(\kappa-a)}+{\cal O}(\epsilon^3)
\right]+{\cal O}(\epsilon^{2\kappa/a})(C-C_\kappa)\ ,
\ee
with
\be
C_\kappa=\frac{\pi\csc(\pi r)}{B(p,1+r)}\ ,
\ee
where $B(a,b)$ is the Euler beta function. We read off from this expansion $\kappa_c=a$ and $C=C_\kappa$, as usual.
As in all previous cases,
the potential is obtained from (\ref{formalV}) and the extremes of the CdL interval come from solving $F(\phi_{0\pm})=0$. 

\begin{figure}[t!]
\begin{center}
\includegraphics[width=0.4\textwidth]{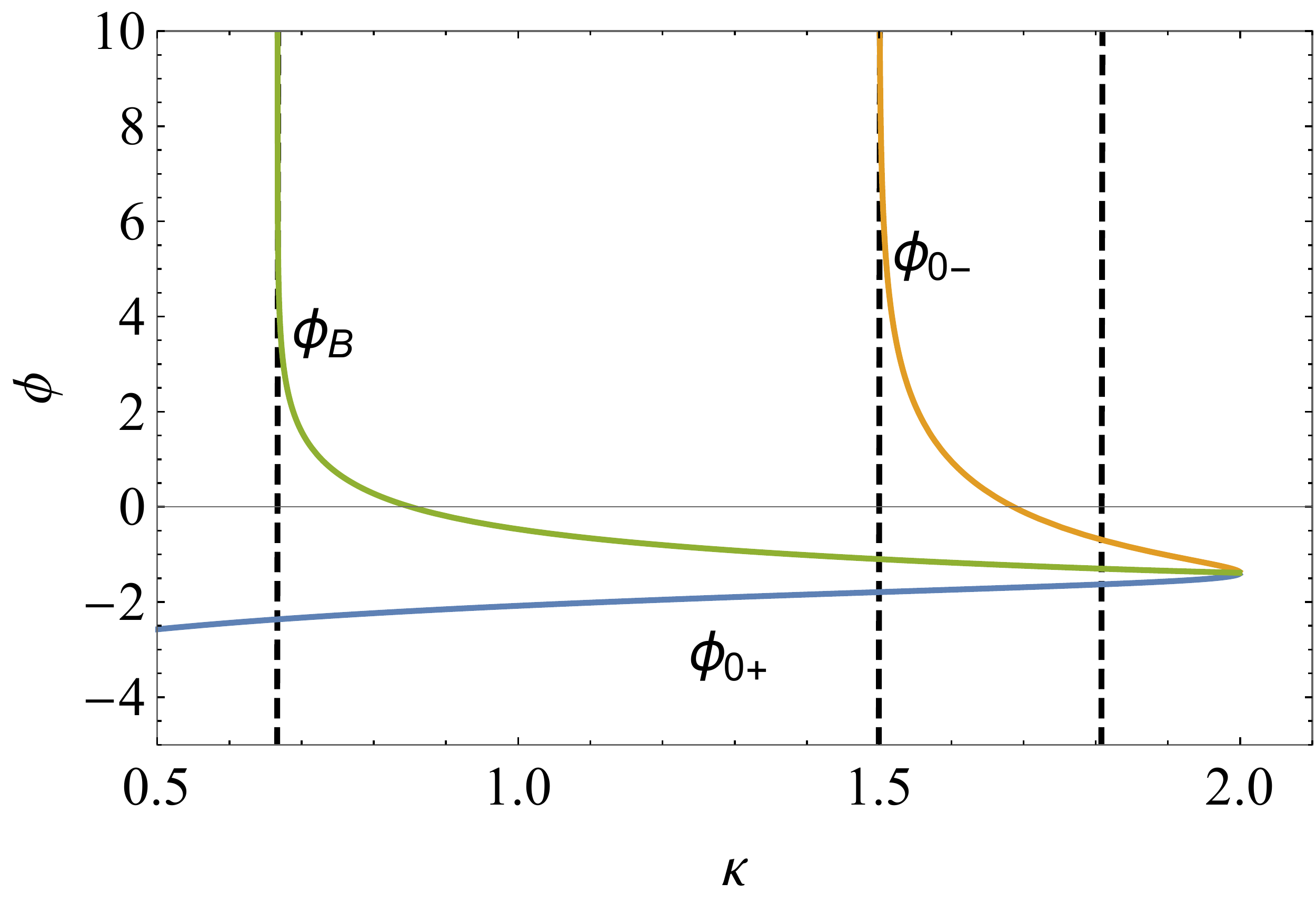}
\includegraphics[width=0.4\textwidth]{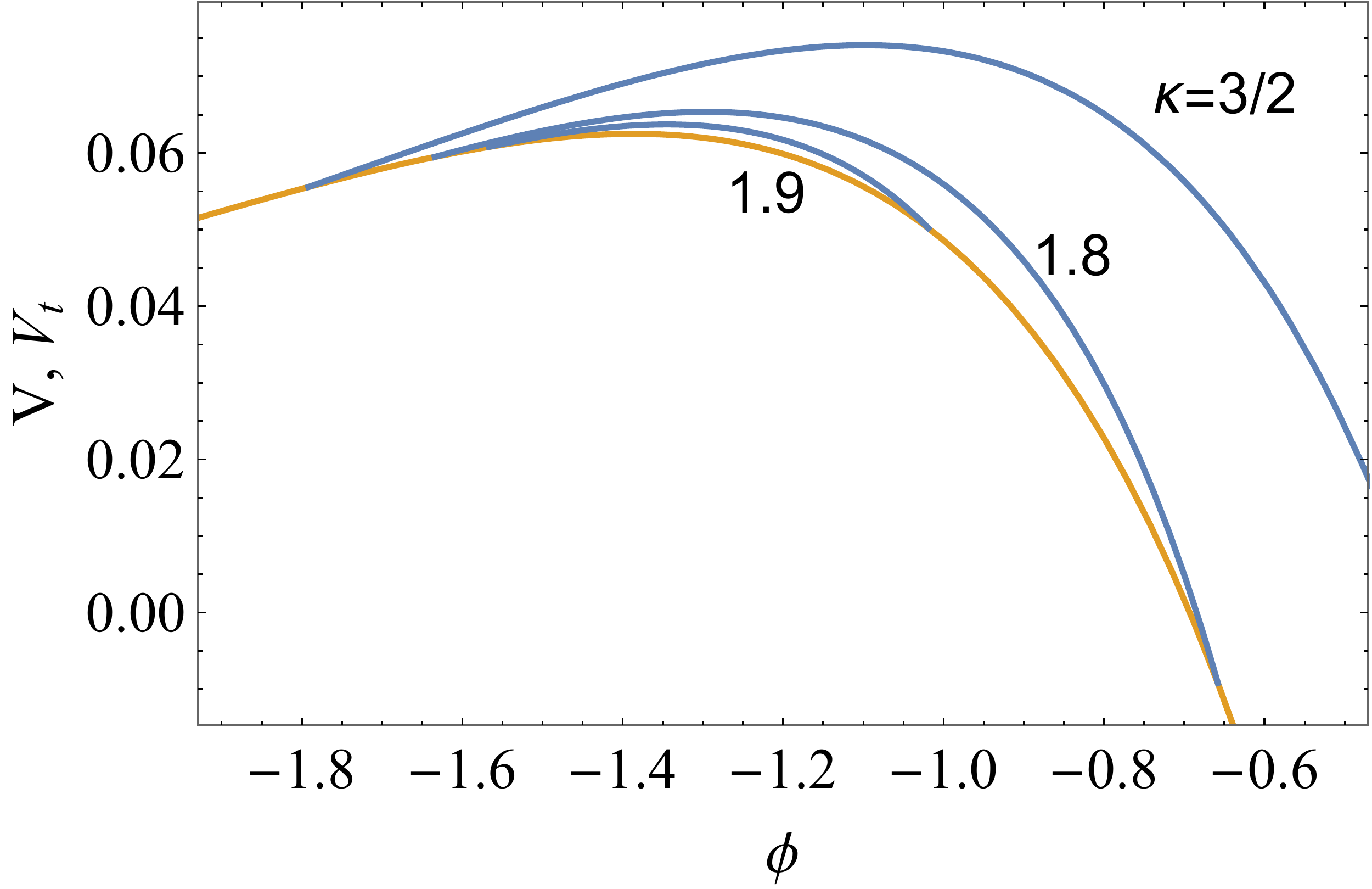}
\end{center}
\vspace{-0.5cm}
\caption{For example \ref{7.5}, varying $\kappa$. Left plot: Limits, $\phi_{0\pm}$, of the CdL interval and $\phi_B$, location of the barrier maximum of $V$. Right plot: For the fixed $V_t$,  several $V$'s in the CdL range for the indicated values of $\kappa$.
\label{fig:75}
}
\end{figure}

For the numerical examples we take $A=1/2$, $a=2$ and leave $\kappa$ free. Fig.~\ref{fig:75}, left plot, shows $\phi_{0\pms}$ as well as the value $\phi_B$ for the barrier top of $V$ as a function of $\kappa$. For $\kappa<2/3$, $V$ grows without reaching a maximum.  For $\kappa<3/2$, $V_t$ never reaches $V$ while for  $\kappa=3/2$, $V$ reaches $V_t$  only asymptotically with $\phi_{0\mm}\to\infty$, corresponding to a CdL field range of infinite extension. Finally, for $\kappa>\kappa_c=2$ there is no CdL instanton. We are therefore interested in principle in the interval $(3/2,2)$ for $\kappa$. 

\begin{figure}[t!]
\begin{center}
\includegraphics[width=0.4\textwidth]{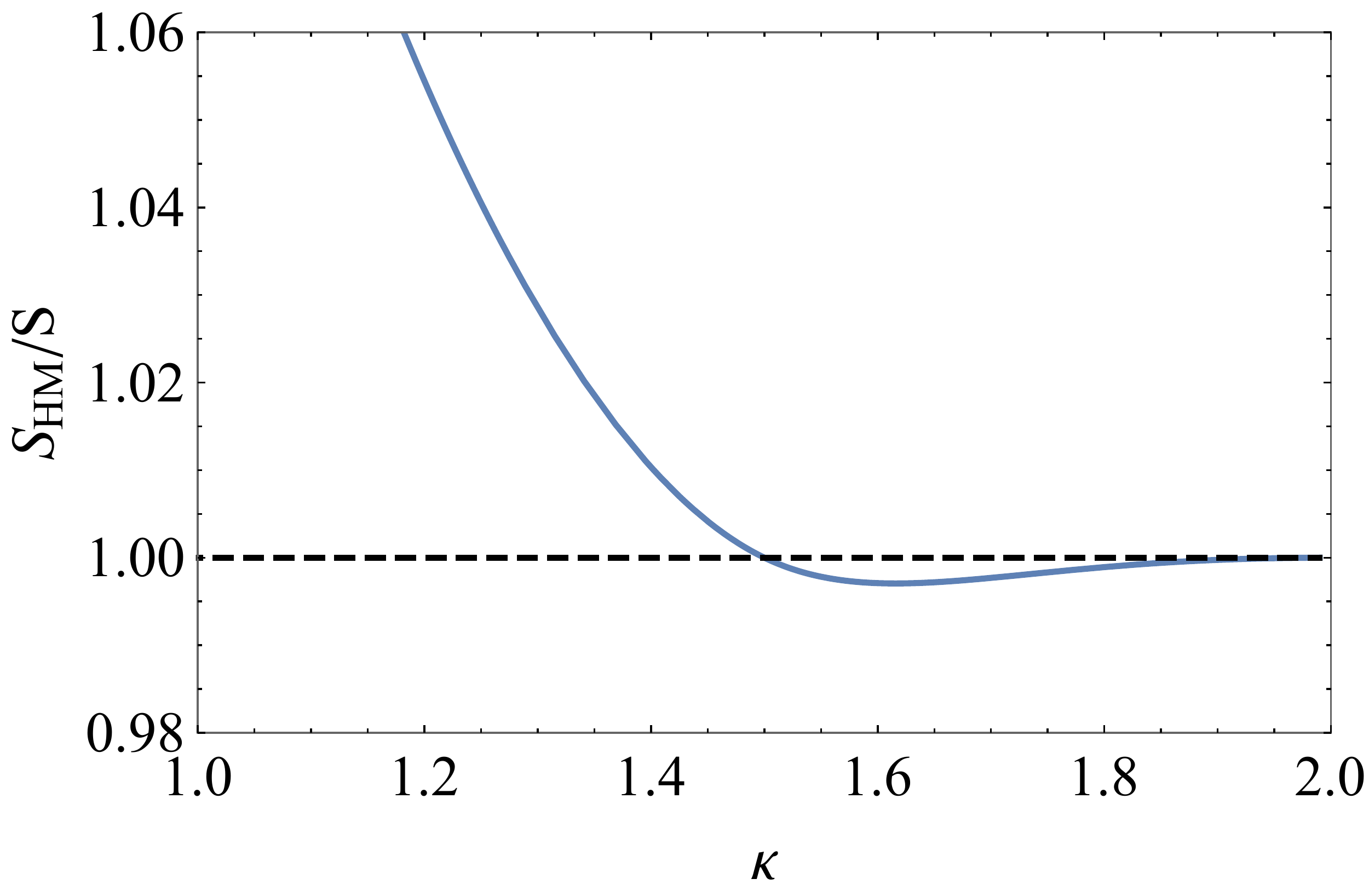}
\includegraphics[width=0.4\textwidth]{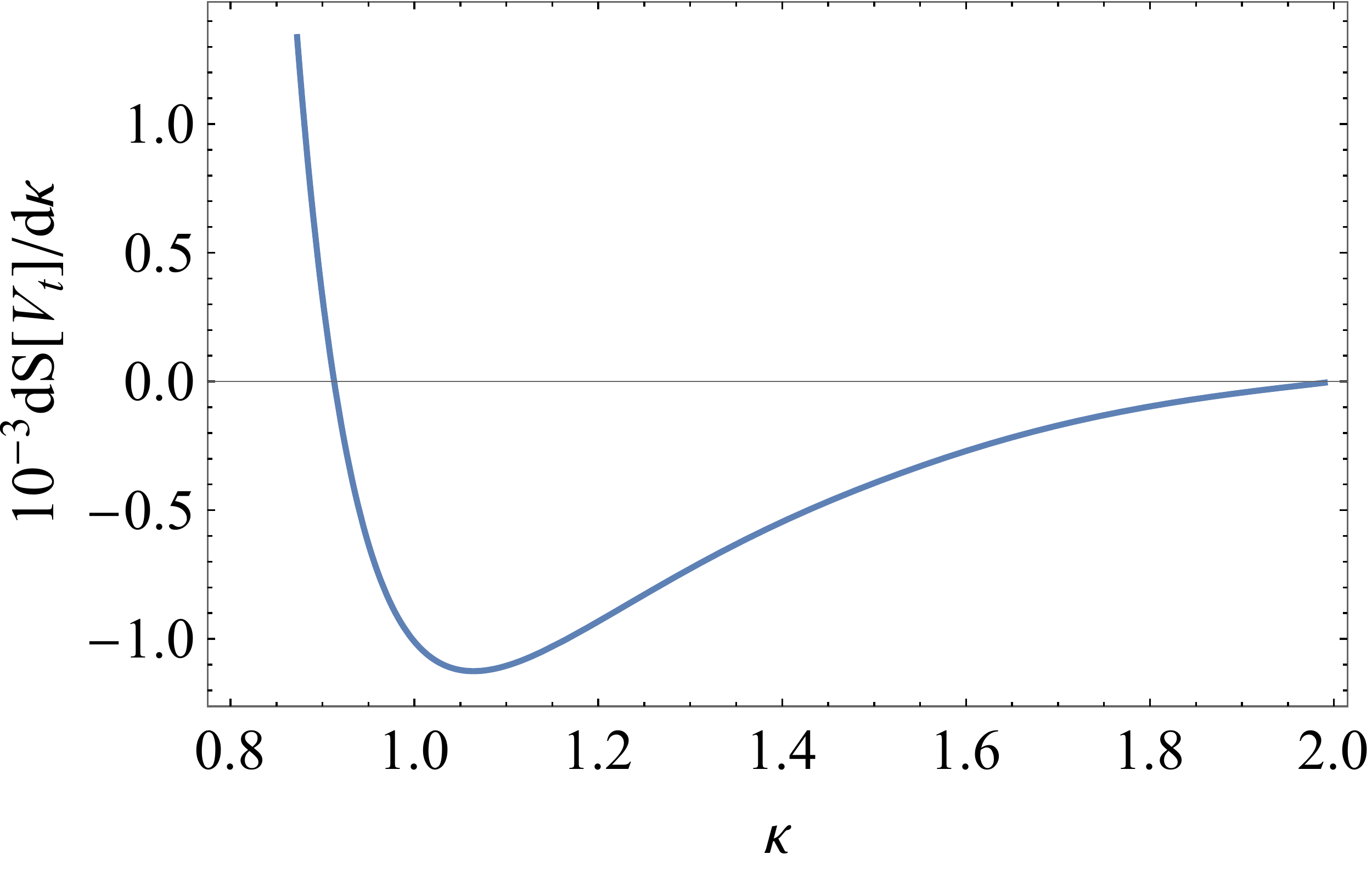}
\end{center}
\vspace{-0.5cm}
\caption{For example \ref{7.5}, ratio of action for decay via Hawking-Moss instanton ($S_{HM}$) to action via Coleman-De Luccia bounce ($S$) as a function of $\kappa$ (left plot). For $\kappa>2$ only the HM transition is possible. Right plot: Slope  $dS/d\kappa$ for the action via the CdL bounce, showing that both signs are possible
\label{fig:actiondSAdS}}
\end{figure}

The tunneling potential, $V_t$, and several potentials, $V$, for different choices of $\kappa$ are plotted in figure~\ref{fig:75}, right plot. We confirm that, as $\kappa\to \kappa_c=2$ the CdL interval shrinks to zero. For $\kappa\simlt 1.81$  the transition is from dS to AdS 
(the left plot also marks by a dashed line this special value).
 For $\kappa=3/2$ the intersection point $\phi_{0\mm}$ of $V$ and $V_t$ goes to $\infty$. This case is discussed below. 

 Figure \ref{fig:actiondSAdS} shows the ratio between the  action for HM decay and the action for decay via the CdL instanton, as a function of $\kappa$. For concreteness we have assumed that the false vacuum 
$\phi_\pp$ sits at a potential height $V_\pp=0.03$ (this just affects the overall value of the actions but not which one dominates). 
The plot shows that decay occurs via a CdL
instanton below $\kappa=3/2$, while for $\kappa>3/2$ the HM instanton has lower action and drives the
decay. Notice also that the CdL part of the action remains finite also below
$\kappa=3/2$ even though the CdL range is infinite in that case (and moreover, $V_t$ never reaches $V$ for $\kappa<3/2$).  This does not seem to be physically accessible for decay, but the action thus obtained might be relevant to bound the true decay action in that range of $\kappa$ (similarly to what happens when a CdL bounce only exists at asymptotic infinity). The situation might be worth looking at in more detail, but goes beyond the goals of this paper. We can also examine  $dS[V_t]/d \kappa$, discussed in section 4.
Performing the integral (\ref{dSdk}) numerically we get the behaviour shown in Fig.~\ref{fig:actiondSAdS}, right plot, with both signs of  $dS[V_t]/d \kappa$ being possible although $dS/d\kappa>0$ occurs only in the suspicious case with $\kappa<3/2$.

The special case $\kappa=3/2$ (still with $A=1/2$ and $a=2$) is particularly interesting. For that critical value of $\kappa$, the potential simplifies dramatically to 
\be
V(\phi) = \frac49 e^{\phi}-\frac23 e^{2\phi}\ .
\ee
Although the CdL field range  is infinite, $(\phi_{0\pp},\phi_{0\mm})=(-\log 6,\infty)$,   the CdL part of the action turns out to be finite, and can be computed exactly: $\Delta S_{CdL}=48\pi^2$. This leads to a total action $S=\Delta S_{HM}+\Delta S_{CdL}=(24\pi^2/\kappa^2)(1/V_\pp-1/V_{0\pp})+48\pi^2$. Remarkably, this is the same as the HM action $S_{HM}=(24\pi^2/\kappa^2)(1/V_\pp-1/V_{B})$. Therefore,  in this case  both instantons coexist and have the same action. Notice that this is very different from previous cases in which the HM case is reached as a limit in which the CdL instanton disappears. It is also interesting that, in the current case, one has  $-3V_B''=4\kappa V_B=4/9$, saturating the usual bound (\ref{mbound}) for the onset of HM. 

In this simple example one can also obtain analytic expressions for the Euclidean bounce and metric profiles, by integrating (\ref{dphi}) and (\ref{rho}),  as
\be
\phi(\xi)=\log\frac{\xi_e^2}{6(2\xi_e-\xi)\xi}\ ,\quad
\rho(\xi)=\frac{\xi(\xi_e-\xi)(2\xi_e-\xi)}{\xi_e^2}\ ,
\ee
with $\xi_e=6\sqrt{6}$. The instanton is compact with the $\xi$ interval being $(0,\xi_e)$. In this case, $\xi_e<\xi_{dS,B}\equiv \pi\sqrt{3/(\kappa V_B)}$, so that the CdL bounce fits comfortably inside the $\phi_B$ horizon.

\section{Summary and conclusions \label{sec:conclusions}}

In this work we have resorted to the tunneling potential ($V_t$) formulation of vacuum decay in quantum field theory, including gravitational corrections \cite{Eg}, to obtain exactly solvable potentials in which to study such decays. The idea  is to find a simple
$V_t$ function (or the auxiliary function $F$ introduced in section~\ref{sec:exact}) whose differential equation of motion can be integrated to get $V$. The method can be used to study decays from any type of false vacua: Minkowski, anti-de Sitter or de Sitter, although we have focused on the latter, which is especially interesting and was lacking in \cite{Eg}. Section~\ref{sec:ex} contains a selection of several examples of such dS decays, including some strikingly simple ones, like the one in subsection \ref{7.4}. The method is quite powerful and many other examples can be found. Previous solvable examples discussed in the literature include some simple cases, like a conformally coupled scalar with a quartic potential with a particular value of the quartic coupling and an AdS false vacuum \cite{exact2} or more general methods based on postulating the Euclidean metric function $\rho(\xi)$, either by deforming the one of  Hawking-Moss instantons \cite{exact1}, or by imposing necessary constraints on the general form of $\rho(\xi)$ \cite{exact0}. Compared with this last more general method, the one we follow has some built-in advantages: for instance, the null-energy condition
$\dot\rho^2-\rho\ddot\rho-1\geq 0$,
imposed by \cite{exact0} (and identified there as the most challenging condition to find a physically meaningful $\rho$) is automatically satisfied by the $V_t$ formulation [as one can check by substituting in the condition above the relations in (\ref{rho})].

As we were particularly interested in finding examples of solvable dS
decays, we first examined how the $V_t$ formulation describes such decays. In particular we showed how these transitions are composed of two different pieces: a Hawking-Moss part (with $V_t=V$) and a Coleman-de Luccia part (with $V_t<V$), with their relative importance set by a balance between their competing needs to be minimized. 
The illuminating discussion of \cite{BW} described a similar split  between a thermal excitation part (corresponding to our HM part) and a tunneling CdL part (corresponding to our $V_t<V$ part).
Our formalism also allows to derive in a straightforward way the simple relation between the actions for two-way transitions between dS vacua  and how, as the energy scale of the potential is raised, the dS decay is dominated by HM instantons. The method also allows
to understand the influence of the shape of the barrier top on this switch from CdL-dominated to HM-dominated decays.
For potentials with a quadratic barrier top we derived in a simple manner the critical value of $V''$ at the top of the barrier below which only the HM persists, showing also how the CdL instanton shrinks away and disappears at that critical value. For potentials with a flatter top the switch from the CdL-dominated decay to the HM-dominated one is qualitatively different, with the CdL instanton coexisting with the HM one although with higher action.

In addition, the general expression for the tunneling action integral allows to obtain several scaling properties of the action for dS decays: (1) Increasing all potential scales by an overall factor (thus increasing the height of the false vacuum but also that of the barrier and of the true vacuum by the same amount) lowers the action, making the potential more unstable; (2) Increasing gravitational effects (e.g. by increasing all mass scales towards $m_P$) lowers the action of dS to dS transitions but can have the opposite effect for dS to AdS transitions, depending on the potential shape and the relative balance of dS and AdS parts of the CdL range; (3)  As one would expect, wider barriers make the decay action
larger, making the vacuum more stable; (4) There is a limit to the effects in (2) and (3) above, given by the relations in (\ref{scalingrange}) which give lower and upper bounds on the action after the indicated rescalings.

\appendix

\section{Conformal Invariant Example}
Without gravity, the scale invariant potential $V(\phi)=-\lambda\phi^4/4$, has a false vacuum at $\phi=0$ that can decay via Fubini-Lipatov \cite{FL} instanton configurations
\be
\phi(r)=\frac{\phi_0}{1+\lambda\phi_0^2r^2/8}\ ,
\ee
where $\phi_0=\phi(0)$ is arbitrary, all having the same tunneling action $S=8\pi^2/(3\lambda)$. In the $V_t$ formulation, 
this family of instantons corresponds to the family of tunneling potentials
\be
V_t(\phi) = -\frac14 \lambda \phi^3\phi_0\ .
\ee

In the presence of gravity, the previous family of instanton field configurations still describes the decay of the false vacuum provided the Lagrangian includes a non-minimal coupling of the field to the Ricci scalar, $\delta {\cal L}=G(\phi)R$, where
\be
G(\phi)\equiv -\frac{1}{2\kappa}+\frac12\xi\phi^2\ ,
\ee
with the conformal value $\xi=1/6$. In the $V_t$ formulation, the tunneling potentials now are
\be
V_t(\phi)= -\frac14\lambda \phi^3\left(\frac{\phi_0-\kappa\phi^3/6}{1-\kappa\phi^2/6}\right)\ .
\ee

The previous discussion is done in the so-called Jordan frame and it is interesting to see how the previous potential and $V_t$ look like in the Einstein frame, after removing the non-minimal coupling by a Weyl rescaling of the metric
\be
g_{\mu\nu}\to \frac{G_E}{G(\phi)}g_{\mu\nu}^E\ ,
\ee
where $G_E=G(0)$ and the index $E$ indicates Einstein frame quantities.
This transformation changes the potentials and kinetic term of the scalar field as
\be
V_E = \left(\frac{G_E}{G}\right)^2V\ ,\quad
V_{tE} = \left(\frac{G_E}{G}\right)^2V_t\ ,\quad
\frac12 \dot\phi_E^2 = \frac{G_E}{G}\frac12 \dot\phi^2\ .
\ee
The latter expression can be integrated exactly, leading to
\be
\phi=\sqrt{6/\kappa}\tanh\left(\sqrt{\kappa/6}\phi_E\right)\ .
\label{canonical}
\ee
Using this relation and the expressions above, one ends up with
\bea
V_E(\phi_E)&=& -\frac{9\lambda}{\kappa^2}\sinh^4\left(\sqrt{\kappa/6}\phi_E\right)\ ,\\
V_{tE}(\phi_E)&=&V_E(\phi_E)-\frac{9\lambda^2}{8\kappa^2}
\sinh^3\left(2\sqrt{\kappa/6}\phi_E\right)\left[\tanh\left(\sqrt{\kappa/6}\phi_{E0}\right)-\tanh\left(\sqrt{\kappa/6}\phi_E\right)
\right]\ ,\nonumber
\eea
where $\phi_{E0}$ is arbitrary and is related to $\phi_0$ in the Jordan frame by (\ref{canonical}).

\section*{Acknowledgments} 
The work of J.R.E. and J.H. has been supported by the Spanish Ministry for Science and Innovation under grant PID2019-110058GB-C22
and the grant SEV-2016-0597 of the Severo Ochoa excellence program of MINECO. The work of J.-F.F. is supported by NSERC.

\end{document}